\documentclass[twocolumn]{aastex631}

\accepted{August 11, 2026}

\submitjournal{APJ}

\usepackage{amsmath}
\begin{document}

\title{Recombination Rates for $5f^N$ Uranium Ions Using the Analogous $4f^4$ Nd III as a Guideline and Their Effect on Kilonova Nebular Spectra}

\correspondingauthor{N. Ferguson}
\email{niamh.ferguson@strath.ac.uk}

\author[0009-0005-3148-513X]{Niamh Ferguson}
\affiliation{Department of Physics, University of Strathclyde, Glasgow G4 0NG, United Kingdom.}

\author[0000-0001-8005-4030]{Anders Jerkstrand}
\affiliation{The Oskar Klein Centre, Department of Astronomy, Stockholm University, AlbaNova, SE-10691 Stockholm, Sweden}

\author[0000-0001-6595-2238]{Smaranika Banerjee}
\affiliation{Astrophysics sub-Department, Department of Physics, University of Oxford, Keble Road, Oxford OX1 3RH, UK}

\author[0000-0002-2160-4546]{M. G. O'Mullane}
\affiliation{Department of Physics, University of Strathclyde, Glasgow G4 0NG, United Kingdom.}

\author{N. R. Badnell}
\affiliation{Department of Physics, University of Strathclyde, Glasgow G4 0NG, United Kingdom.}

\begin{abstract}
Dielectronic recombination (DR) is expected to be the dominant recombination process during the non-local thermodynamic equilibrium (non-LTE) phase of kilonovae, yet reliable DR data remain unavailable for most heavy ions. 
Current spectral models therefore rely on simplified recombination prescriptions, introducing significant uncertainties into predicted spectra.
We present an optimization strategy for open f-shell ions using \texttt{AUTOSTRUCTURE}, targeting uranium ions U II--U IV relevant to actinide-producing kilonovae. 
As a benchmark case, calculations are performed for Nd III to validate the treatment of the f-shell structure and its impact on DR. 
The resulting DR rate coefficients are of order $10^{-10}$--$10^{-12}$ cm$^{3}$\,s$^{-1}$ over temperatures relevant to kilonova plasmas. 
The optimized rates are implemented into the radiative-transfer code \texttt{SUMO}, and their impact on kilonova spectra  studied. 
The Nd III benchmark demonstrates that refinements to the atomic structure can produce measurable changes in spectral features supporting the atomic structure optimisation performed for U II--IV.

\end{abstract}

\section{Introduction} \label{sec:intro}

The observation of the kilonova associated with the binary neutron star merger GW170817 provided the first direct evidence that neutron star mergers are a major site of rapid neutron-capture (r-process) nucleosynthesis, helping to resolve the long-standing question of the origin of the heaviest elements in the Universe.
Spectroscopic observations can aid the identification of elements produced in kilonovae, but only a few have been proposed so far, with complex physics, line blending, and limitations on atomic data making spectra challenging to analyze. Research into the atomic properties \citep[e.g][]{DK2017,FontesLa,FontesAc} and radiative transfer modelling \citep[e.g.][]{Tanaka2013,Wollaeger2018} have mainly focused on the early phases, up to about a week post-merger. 
Various charged heavy ions, up to 10 times ionized in the first day post-merger, are produced and may be detected at early times  \citep{banerjee_simulations_2020}.  
Over time, the expanding ejecta recombines and eventually reaches a low enough density that a local thermal equilibrium (LTE) can no longer hold. In the ensuing non-LTE (NLTE) phase, level populations depend on many collisional and radiative rates \citep[e.g.,][]{Jerkstrand2025}. Ionization balance is the prime motivator for calculating DR rate coefficients for the expected elements in the NLTE phase of kilonovae.

In regards to atomic data, lanthanide and actinide data have been produced for the early time post-merger by various groups to aid opacity calculations with low charged data produced \citep{DKNB,DK2017,Tanaka2020,FontesLa,FontesAc,Deprince_Wagle_Nasr_Gallego_Godefroid_Goriely_Just_Palmeri_Quinet_Eck_2025} and also higher ionization stages presented \citep{banerjee_simulations_2020, Carvajal_Gallego_Berengut_Palmeri_Quinet_2021, Banerjee, CarvajalGallego_Deprince_Berengut_Palmeri_Quinet_2023, CarvajalGallego_Deprince_Palmeri_Quinet_2023, banerjee_diversity_2024, Gallego_Deprince_Maison_Palmeri_Quinet_2024}. 
In the NLTE phase, the conditions of the ejecta involve low/moderate temperatures, $\lesssim10^4$\ K, and low density $\lesssim 10^{8}$\ cm$^{-3}$ \citep{Hotokezaka}. It is only quite recently that research groups have turned their focus to the NLTE phase of kilonovae \citep{Hotokezaka,PognanNLTE} with variations of models and atomic data of different processes for the phase in the literature \citep{Gillanders_Smartt_2025,McCann_Ballance_McNeill_Sim_Ramsbottom_2025,banerjee25,Mulholland_Ramsbottom_Ballance_Sneppen_Sim_2026,Jerkstrand2026,Deprince_Maison_2026,FerreiradaSilva2026}.

Calculation of recombination rates of Se and Kr have earlier been carried out by \citet{Sterling2011-Se} and \citet{Sterling2011-Kr}, for application to planetary nebulae but equally applicable to the late-stage modelling of kilonova.
Work has so been carried out to investigate the recombination rates for some ion stages of Nd \citep{Hotokezaka} and various ionization stages of Se, Rb, Sr, Y, and Zr \citep{banerjee25} using \texttt{HULLAC} and Y$^+$, Sr$^+$, Te$^2+$ and Ce$^2+$ presented by \cite{Singh_Harman_Keitel_2025}. 


As there are likely regions of different electron fraction $Y_e$ in kilonova ejecta, with only some producing lanthanides and/or actinides, it is of interest to investigate recombination rates for both a lanthanide representative, such as neodymium (Nd), and an actinide representative, such as uranium (U). A model including a significant actinide abundances would have an electron fraction $Y_e \lesssim 0.15$ \citep{Wanajo_2018,PognanNLTE}, and the actinide abundance also varies significantly between nuclear networks \citep{Pognan2025}. Although the actinide abundances are still relatively low also for low-$Y_e$ material, they may give spectral signatures \citep{Pognan2025}. 

The paper is structured as follows. The background theory of recombination is discussed in section \ref{sec:theory}. The details of the atomic data applied to neodymium and uranium is discussed in section \ref{sec:application}. The recombination calculation results and influence on spectra are presented in section \ref{sec:results}. Finally, we present discussion and conclusions in section \ref{sec:Discussion}.

\section{Theory} \label{sec:theory}
In NLTE, dielectronic recombination (DR) is often the dominant electron-ion recombination process for many laboratory and astrophysical plasmas \citep{Burgess_1964,FeDR}. This arises because DR proceeds through resonant electron capture which can significantly enhance the recombination cross section compared with non-resonant radiative recombination, particularly under conditions where the relevant autoionizing states are efficiently populated.
DR \citep{Burgess_1964} is a two-step process involving the capture of a continuum electron, resulting in an ion core excitation with subsequent radiative stabilisation of the core: 
\begin{equation}
    A^{q+}+e^- \leftrightarrow A^{**(q-1)+} \rightarrow A^{(q-1)+} + \gamma.
\end{equation}
The captured electron will occupy a highly excited level of the ion, and the stabilisation of the ion can happen in one of two ways.
The captured electron can be ejected whilst the ion core electron relaxes into the initial ground state, known as autoionization.
If the doubly excited state $(A^{**(q-1)+})$ radiates instead of autoionizing, a photon ($\gamma$) is emitted such that the final state is bound.

The DR rate coefficient \citep{DR} can be calculated according to

\begin{equation}\label{eq:DRRate}
\begin{split}
    \alpha_{DR}(i+e^- &\rightarrow f) = {\left({\frac{4\pi{\alpha_0}^2 I_H}{kT}}\right)}^{\frac{3}{2}} \sum_{j}\frac{g_j}{2g_i}e^{{-E}/{kT}} \\
    & \times \frac{\sum_l A_a(j\rightarrow i,kl) A_r(j \rightarrow f)}{\sum_h A_r(j\rightarrow h) + \sum_{m,l} A_a(j\rightarrow m,kl)},
\end{split}
\end{equation}
where $i$ and $f$ are the initial and final states, $j$ is the doubly excited intermediate state, $E$ is the energy of the continuum electron fixed by the position of the resonances, $g_i$ is the statistical weight of the \textit{N}-electron target ion, $g_j$ is the statistical weight of the (\textit{N}+1)-electron intermediate state and $T$ is the temperature (K). Equation \ref{eq:DRRate} is the sum over all true bound states, $f$, $j$ over all intermediate states formed from dielectronic capture from initial state, $i$.

The dielectronic capture rate coefficient, ${{(\frac{4\pi{\alpha_0}^2 I_H}{kT})}^{\frac{3}{2}}\sum_{j}\frac{g_j}{2g_i}e^{{-E}/{kT}} \times \sum_l A_a(j\rightarrow i,kl)}$, is multiplied by the fraction that stabilise, ${A_r(j \rightarrow f)}/ {\sum_h A_r(j\rightarrow h) + \sum_{m,l} A_a(j\rightarrow m,kl)}$ as shown in equation \ref{eq:DRRate}.

Autoionizing states just above threshold contribute to the low-temperature DR and it is illustrated in equation \ref{eq:DRRate} for any given temperature, the exponential factor shows that the energies need to be within a low-temperature range.

The high-temperature DR peak is mainly produced from core radiation stabilisation, which is useful for applications to hot astrophysical plasmas such as the solar corona. 
For kilonova applications, however, the low-temperature DR is of primary importance, as nebular-phase ejecta are expected to have temperatures of only a few thousand kelvin (up to $\sim 10^5$K depending on epoch and ejecta conditions, (eg. \cite{Hotokezaka, Pognan22a,Pognan2025,banerjee25}).

Another recombination process to consider is radiative recombination.
\begin{equation}\label{eq:RR}
    A^{q+}+e^- \rightarrow A^{(q-1)+}+ \gamma.
\end{equation}
Radiative recombination (RR) \citep{RR} is the process of an electron being captured by a positively charged ion and emitting a photon simultaneously, such that the final state is bound, as shown in equation \ref{eq:RR}.

It can also be useful to compare the RR rate coefficients with the DR rate coefficients, to emphasise that DR can dominate the RR significantly in some instances. 
Work conducted by \cite{RR} discusses the method of calculating RR cross sections using \texttt{AUTOSTRUCTURE}, as is used in this work.
The process of calculating the RR rate coefficients is more straightforward than that of the DR rate coefficients and is discussed in detail in \cite{RR}.

\section{Application of Atomic Data to Kilonovae} \label{sec:application}
The atomic structure of an ion is important when obtaining recombination rates.
\texttt{AUTOSTRUCTURE} \citep{AS} was used to calculate the atomic data necessary for recombination rate calculations by producing intermediate coupling data files and structure information.

Specifically for lanthanide and actinide ions, the $nf \rightarrow nf$, $nf \rightarrow nd$ and the $nf \rightarrow ns$ transitions are important for the relevant temperature range for kilonovae, where $n$ is the principal quantum number, which is 4 or 5 for the $f$-shell for the lanthanide and actinides, respectively. 
The reduced model configuration sets are as given in table \ref{tab:configs}.
\begin{table*}[ht]
\centering
\caption{The master configuration sets and N+1 rydberg configurations used within the recombination calculations.}
\label{tab:configs}
\renewcommand{\arraystretch}{1.2}
\begin{tabular}{lcc}
\hline
Ion & Configuration & N+1 Rydberg configurations \\
\hline
U II & $5f^3\{7s^2 , 6d7s\}$ & $5f^3\{6d7s^2,6d^27s\}n\ell$ \\
U III & $5f^4, 5f^3\{6d,6s\}$ & $5f^5 n\ell, 5f^46d n\ell, 5f^3\{6d^2,7s^2\}n\ell$ \\
U IV & $5f^3, 5f^2 \{6d,6f,6g\}, 5f\{6d^2,6f^2,6g^2\}$ & $5f^4 n\ell,5f^3\{6d,6f,6g\}n\ell, 5f^2\{6d^2,6f^2,6g^2\}n\ell$ \\
U V & $5f^2, 5f\{6d,6f,6g\}, 6d^2$  & $5f^3n\ell, 5f^2\{6d,6f,6g\}n\ell, 5d6d^2n\ell$ \\
Nd III & $4f^4, 4f\{5d,6s\}$  & $4f^5 n\ell, 4f^35d^2 n\ell,4f^45d n\ell$ \\
\hline
\end{tabular}
\end{table*}
A reduced set was deduced from an initial configuration-averaged calculation which was used to determine the dominant contributions to recombination over the temperature range of interest. Some higher energy-lying configurations that were found from preliminary configuration-average calculations to make negligible contributions to the total recombination rate over the temperature range considered in this work were omitted. 
This reduction significantly decreases the computational expense while retaining the dominant physical processes governing the recombination.
In the computation of DR, a Rydberg electron was coupled to each target configuration with explicit calculations performed for a maximum $n=35$ and $\ell=15$ and intermediate values up to $n=999$.

Before the DR rate coefficients were calculated, the structure of the ion was investigated through the positioning of the energy levels. The atomic structure was very sensitive to the choice of theoretical description. 
\texttt{AUTOSTRUCTURE} solves the total wavefunction in a Thomas-Fermi potential, which contains an orbital scaling parameter $\lambda$. 
It was found that small alterations to the scaling parameter on the $f$-shell resulted in large shifts in the atomic structure. 
By changing the scaling parameters, a cascade of changes occurs. Altering the scaling parameters changes the wavefunctions used within the \texttt{AUTOSTRUCTURE} calculations, and this inherently changes the diagonalised Hamiltonian, which produces a change in eigenvalues.

An initial out-of-box structure run was done to check where energy levels were positioned. This structure simply took default values used within \texttt{AUTOSTRUCTURE}.
The DR rate coefficient was calculated for this out-of-box calculation to provide a comparison with the optimised DR.
The structure of the ion can be adjusted to correspond with some observed levels, and one can argue that this atomic structure is more accurate than without adjustment via $\lambda$.
Observed energies from the NIST levels database \citep{NISTNd,Brewer,BlaiseWyart,Ding_Ryabtsev_Kononov_Ryabchikova_Clear_Concepcion_Pickering_2024} were used as guide to adjust the theoretical energies using the optimising parameter, $\lambda$, to obtain a more realistic atomic structure and to confirm the correct ground level.
Making adjustments via the scaling parameter allows the configurations to be moved energetically so that the near ground levels can be brought close to experimental data and the subsequent levels will follow.
Theoretical low charge DR calculations can illustrate good levels of agreement with experimental values \citep{Experimental}.
Due to this, this method of obtaining the atomic structure can be considered a suitable method when observed energies are available.
A large configuration expansion was not adopted in the present calculations, despite the potential for improved atomic-structure accuracy. 
The primary consideration was the computational expense of the subsequent recombination calculations, which are particularly demanding for actinide ions in terms of both CPU time and data storage requirements. 
Expanding the configuration set substantially would increase the complexity and size of the calculation to a level that would make the recombination modelling prohibitively expensive. 
Consequently, a more restricted configuration basis was employed to provide a balance between an adequate description of the target ion and the practical requirements of generating recombination data suitable for astrophysical applications. 
Furthermore, increasingly extensive configuration expansions do not necessarily yield significant improvements in the low-temperature recombination rate coefficients of interest, particularly where the rates are sensitive to the positions and mixing of near-threshold resonances \citep{Sterling2011-Kr,Badnell_Ballance_Griffin_OMullane_2012}.
The influence of configuration expansion and its convergence has also been investigated in the context of opacity calculations for kilonova modelling \citep{Flors_Silva_Deprince_CarvajalGallego_Leck_Shingles_Martinez-Pinedo_Sampaio_Amaro_Marques_etal._2023}. 

Should the splitting for the calculated data be larger than that of the observed, introducing the observed energies allows the calculated splitting to be altered to the smaller ones, ensuring that the resonances are positioned correctly. 
Observed energies correct the position of the resonances which affects the low-temperature DR as the resonances will move closer/further to the threshold and increase/decrease the DR rate coefficient. Adjusting the resonances ensures the DR rate coefficient is more accurate since the structure now includes experimental validation.

After obtaining the reliable structure for an ion, the DR cross-sections and rate coefficients can be calculated.
For the neodymium (Z=60) ion, Nd III, the NIST \citep{NISTNd} database and energies from \cite{Brewer,Ding_Ryabtsev_Kononov_Ryabchikova_Clear_Concepcion_Pickering_2024} were used as a benchmark for the term's energy positions, using only the $4f$ orbital scaling parameter, $\lambda$, to adjust these into a more accurate position in terms of energy and validate our low energy atomic structure.
The initial investigation relied solely on the optimisation of the $4f$ orbital and for the low-lying energy levels to be applicable to the low-temperature regime of interest.
It was expected the structure would be highly sensitive to this orbital, as was found to be true especially for the low-lying energy levels.
Optimisation of the other orbitals becomes of importance to obtain the high temperature DR features. 

For the Uranium (Z=92) ions, the data available on NIST ASD was limited to only the ground term, so the energies from work by \cite{BlaiseWyart} were used as they expand upon energies obtained from \cite{Brewer}.
Optimising the structure pre-diagonalisation of the Hamiltonian allows the potentials to be appropriately adjusted which results in a better structure of the wavefunctions.
The benchmarking case of the Nd III ion, optimising only the $f$-shell orbital, provided a scheme of obtaining a good low-energy structure that could be used to model the low-temperature regimes of complex actinide ions.

In the DR calculation, continuum electrons and Rydberg electrons are coupled to the target configurations used within the calculation.
The N+1 configurations were also included in the DR calculations to allow for outer electron stabilisation.

\section{Results} \label{sec:results}
\subsection{Recombination} \label{sec:Recombination}
Dielectronic recombination (DR) dominates at low temperatures for any ion with fine structure levels within the ground term.
Recombination rates are presented for Nd III as a source of benchmarking our method of low-temperature optimisation to compare with \cite{Hotokezaka}.
Recombination rates are presented for uranium ions II-V to investigate their impact on spectral models and determine whether one should expect uranium to contribute significantly enough to warrant further study.

\begin{figure*}[ht]
    \centering
    \includegraphics[width=0.5\linewidth]{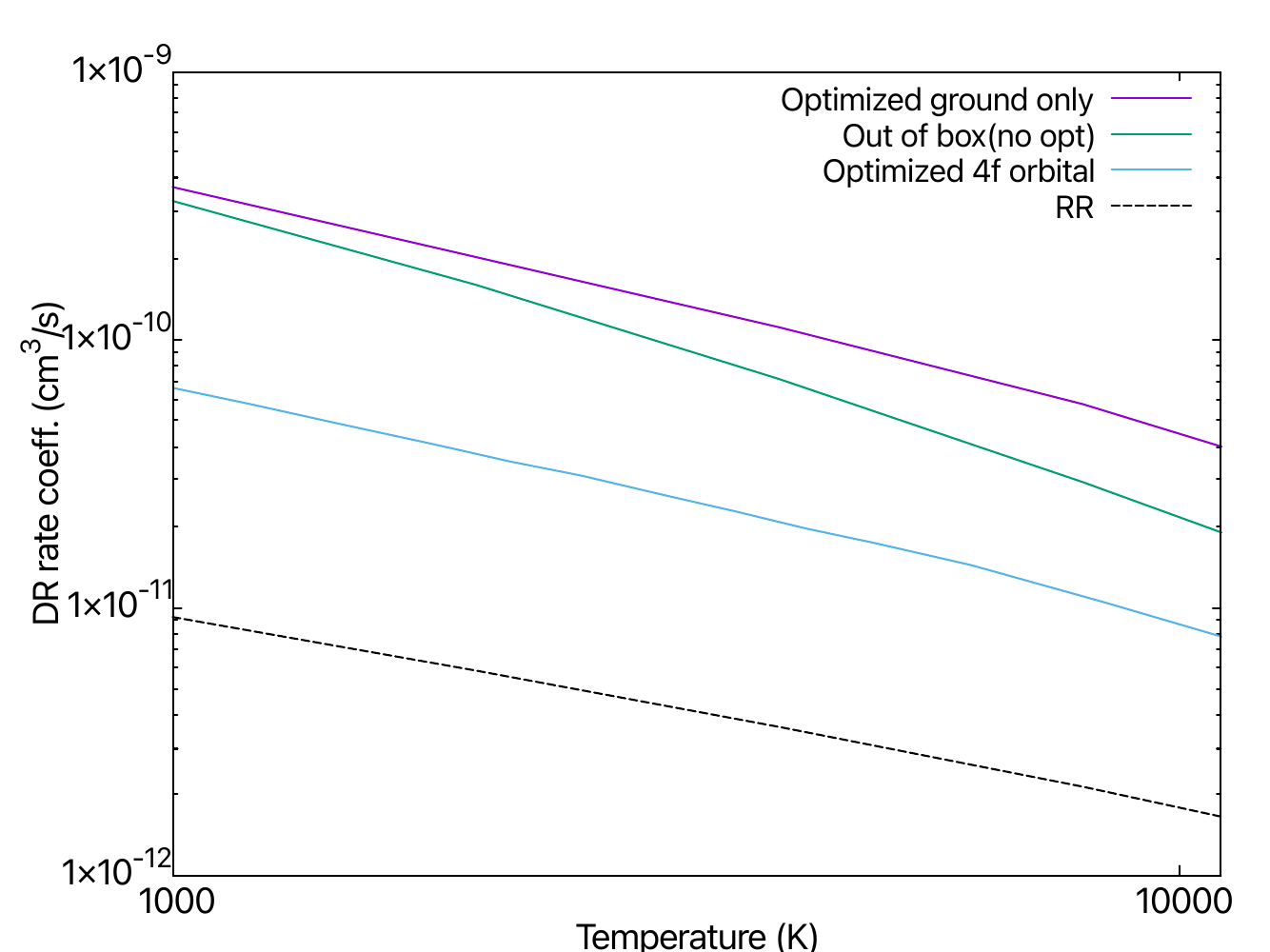}
    \includegraphics[width=0.4\linewidth]{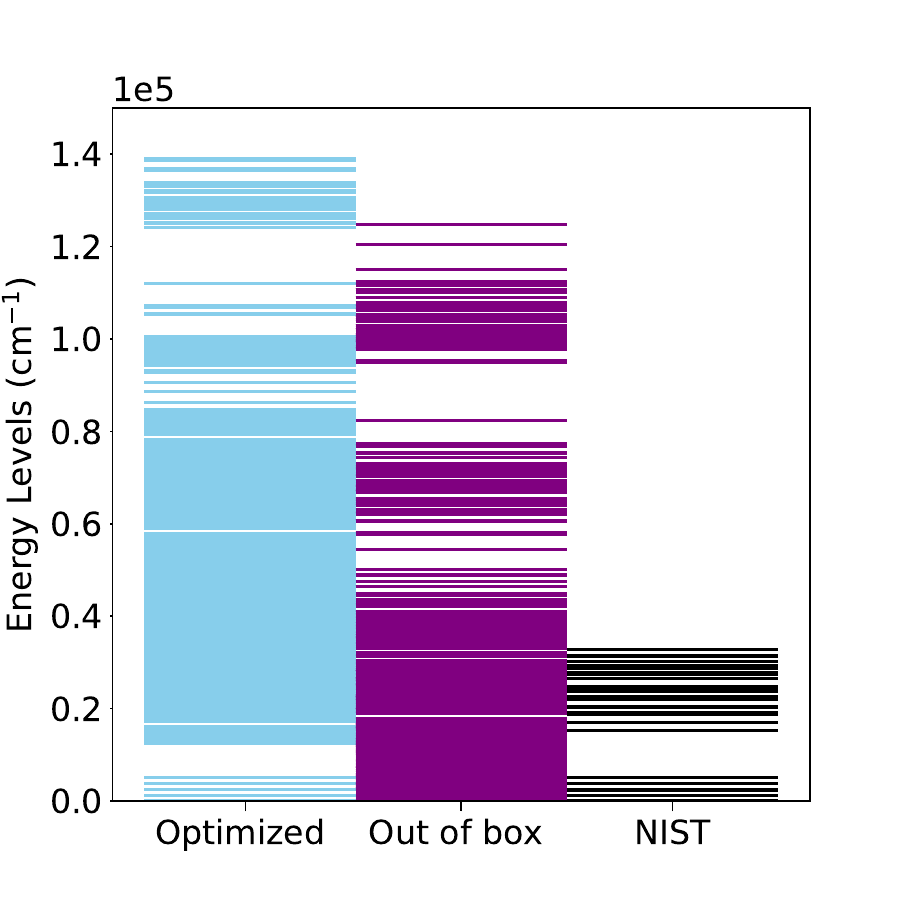}
    \caption{Left: DR rate coefficient of Nd III $\rightarrow$ Nd II using the out of the box structure (green, with incorrect ground and configuration ordering), the optimised 4f shell structure (light blue) and minimal optimization to only obtain the correct ground but neglecting the other level positions (purple). The RR is produced from the structure corresponding to the optimized 4f orbital. Right: The energy levels of the two optimized structures and NIST available data.}
    \label{fig:NdObsCalc}
\end{figure*}

From figure \ref{fig:NdObsCalc}, one can see that the atomic structure can alter the DR rate coefficients calculated and the impact of including observed energies on total DR can be investigated.

The default out-of-box structure calculation for Nd III was poor.
The ground configuration was incorrect and energy levels were at vastly different positions compared to the available experimental data and the configuration ordering was also incorrect for this structure. 
A comparison of the first 5 levels from the $4f^4$ configuration due to the different structures is presented in the table \ref{tab:levels}. 
\begin{table}[ht]
\centering
\caption{Comparison of the first five energy levels of the $4f^4$ configuration}
\label{tab:levels}
\begin{tabular}{lccc}
\hline
Level & NIST (Ry) & Out-of-box (Ry) & $4f$ orbital\\
 & &  & optimised (Ry)\\
\hline
$^5I_4$ & 0.000000 & 0.86584 & 0.00000\\
$^5I_5$ & 0.010368 & 0.86658 & 0.00739\\
$^5I_6$ & 0.021757 & 0.86746 & 0.015639\\
$^5I_7$ & 0.033853 & 0.86845 & 0.024492\\
$^5I_8$ & 0.046414 & 0.86956 & 0.033772\\
\hline
\end{tabular}
\end{table}
The ground configuration for the out-of-box structure was $4f^35d$ where a small optimization was done to obtain the correct ground only to explore the DR comparisons in figure \ref{fig:NdObsCalc}.

The DR rate coefficient was computed for this out-of-box structure and is shown with the green line in figure \ref{fig:NdObsCalc}. 
Minimal effort was made to only ensure the correct ground configuration and term but no effort was made for the other levels to investigate the DR rate coefficient for this poor structure of Nd III with only a correct ground term.
The DR rate coefficient was found to be very similar to the initial, non-optimised, out-of-box run.

The structure was then optimised to ensure an overall better structure, ensuring that, at least, the lowest-lying level of each configuration was at the appropriate energy position using data from \cite{Brewer}, \cite{BlaiseWyart} and \cite{Ding_Ryabtsev_Kononov_Ryabchikova_Clear_Concepcion_Pickering_2024}.
The structure was optimised by adjusting $\lambda$ scaling parameters on the valence orbitals.
Small adjustments from the unity scaling parameters caused large changes in the structure,  emphasising the sensitivity of the open $f$-shell. The scaling parameter was adjusted so the low-lying energy levels matched those available from \cite{NISTNd}. 
The NIST data was limited but contained some levels up to 0.27 Rydberg. 

The optimised structure was used to calculate the corresponding DR rate coefficient, shown in figure \ref{fig:NdObsCalc}. It can be seen that the optimised DR rate coefficient is smaller than that of the out-of-box (OOTB) structure and the structure with only the ground level optimized. This is due to the rate produced from these poor structures having many resonances near threshold which should not be as compact or as near threshold. 
The optimised DR rate coefficient has resonances near threshold but not as compact or close compared to the poor structure. These near-threshold resonances drive the DR rate coefficient higher in value and cause sensitivity to the DR rate coefficient. 
A further adjustment of the 4f orbital was done to ensure energy level positions up to 0.8 Ry were in an appropriate position. This further adjustment slightly spread the low-lying energy levels which agreed with NIST and put low-lying resonances in the correct position.

The DR rate can be extended to higher temperatures for other applications, where one would see a high-temperature peak, but here is restricted to the low temperature range for kilonovae study.
 
However, further investigation into the other Nd ions and high temperature features is in progress due to complexities and sensitivities in the structure which arose during this work and will be published in an upcoming work.

\begin{figure}[tb]
    \centering
    \includegraphics[width=1\linewidth]{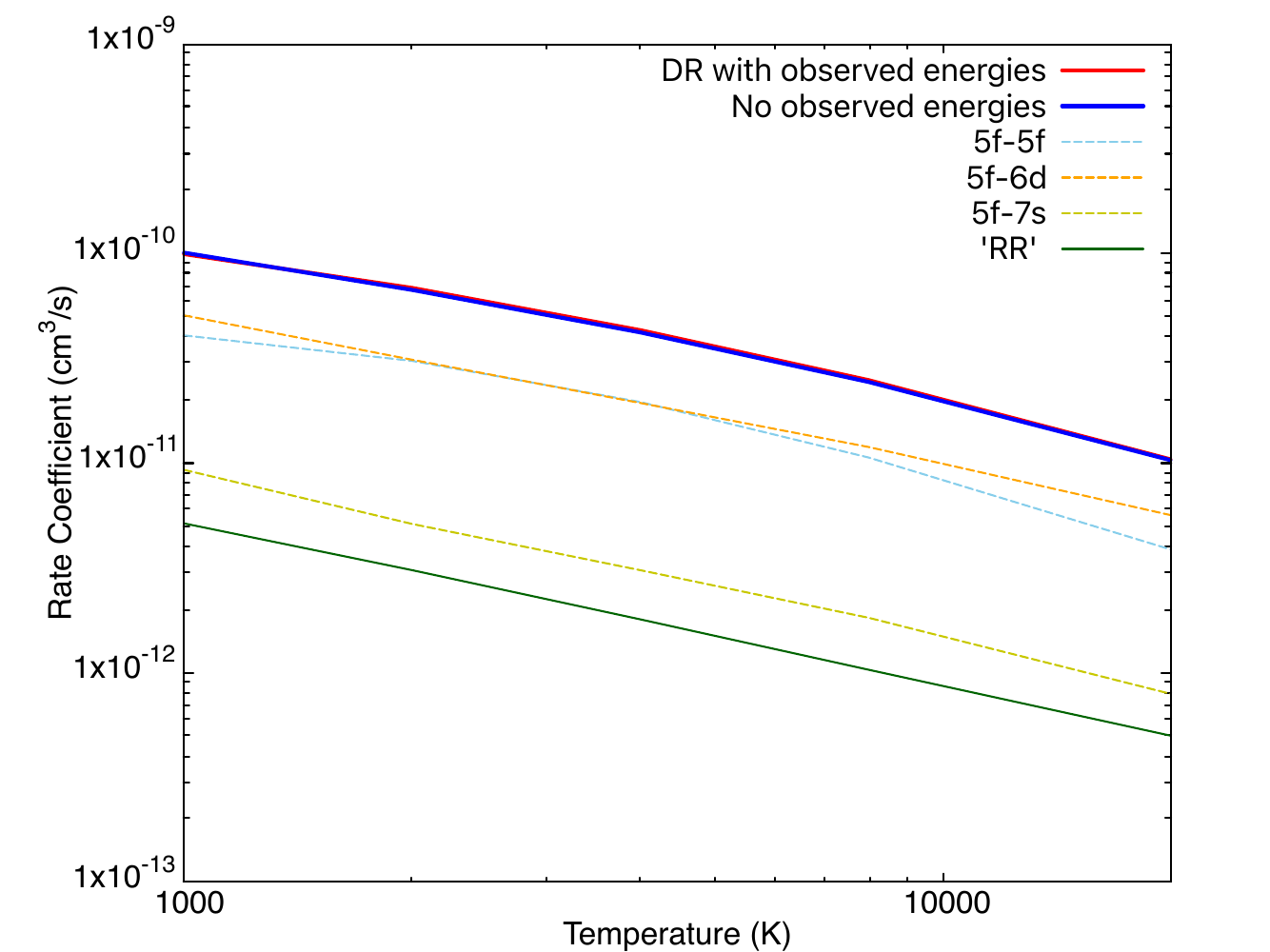}
    \caption{DR and RR rate coefficients of U III $\rightarrow$ U II.
    The DR rate coefficients, including observed energies (red line) and excluding observed energies (blue line), are shown.
    Individual core transition contributions are highlighted to show that the DR rate coefficient for U III is dominated by the $5f\rightarrow 5f$ transition (dashed blue line) and the $5f\rightarrow 6d$ transition (dashed orange line).
    The $5f\rightarrow 7s$ transition does not contribute much to DR, as illustrated by the dashed yellow line.}
    \label{fig:UDR}
\end{figure}

Due to having a similar structure to Nd III, U III was initially computed.
From the initial structure run of U III, the energies of the ground terms were positioned much higher than that of the observed energies from \cite{Brewer} and \cite{BlaiseWyart}.
As with the Nd III case, the energies were adjusted by optimising on the $f$-shell only using the $5f$ orbital scaling parameter to adjust energy levels in agreement with the data from \cite{BlaiseWyart}.
The calculated splitting for the ground terms of U III were similar to the observed after this $5f$ scaling adjustment, resulting in the resonances staying in a similar position with respect to the threshold after the inclusion of the observed energies in the post processing shift stage.
This is shown in figure \ref{fig:UDR}, where the red and blue line denoting the post-processing shift to external data agree with each other, confirming the adjusted structure's accuracy against that of \cite{Brewer} and \cite{BlaiseWyart}.

The DR rate coefficient for U III $\rightarrow$ U II, shown in figure \ref{fig:UDR}, demonstrates individual transition contributions to the total DR rate coefficient.
Including observed energies (red line obscured by blue) in the post-processing stage resulted in a DR rate coefficient of $\alpha_{DR}\sim4.29\times10^{-11}$\ cm$^3$s$^{-1}$ at $T=4000$ K, remaining similar to the DR rate without the observed energies (blue line).
This is due to obtaining an accurate optimised atomic structure with energy levels that closely agreed with the \cite{BlaiseWyart} data.

The transition $5f \rightarrow 7s$, seen in figure \ref{fig:UDR} (yellow line), does not contribute much to the total DR with a DR rate coefficient of $\sim 3.07\times10^{-12}$\ cm$^3$s$^{-1}$ at $T=4000$ K.
However, it still clearly dominates the RR rate coefficient by a factor of 4. 
Whereas the $5f \rightarrow 5f$ and $5f \rightarrow 6d$ transitions contribute the most to the DR rate coefficients with $\alpha_{DR}\sim2\times10^{-11}$\ cm$^3$s$^{-1}$ at $T=4000$ K.
The low-temperature DR rate coefficient for U III $\rightarrow$ U II at $T=1000$ K was calculated to be $\alpha_{DR}\sim10^{-10}$\ cm$^3$s$^{-1}$, an order of magnitude higher than the canonical rate of $10^{-11}$ cm$^3$s$^{-1}$ currently being used for modelling.
The RR rate was calculated to be $\alpha_{RR} \sim 2\times10^{-12}$\ cm$^3$s$^{-1}$ at $T=4000$ K.

\begin{figure}[ht]
    \centering
    \includegraphics[width=1\linewidth]{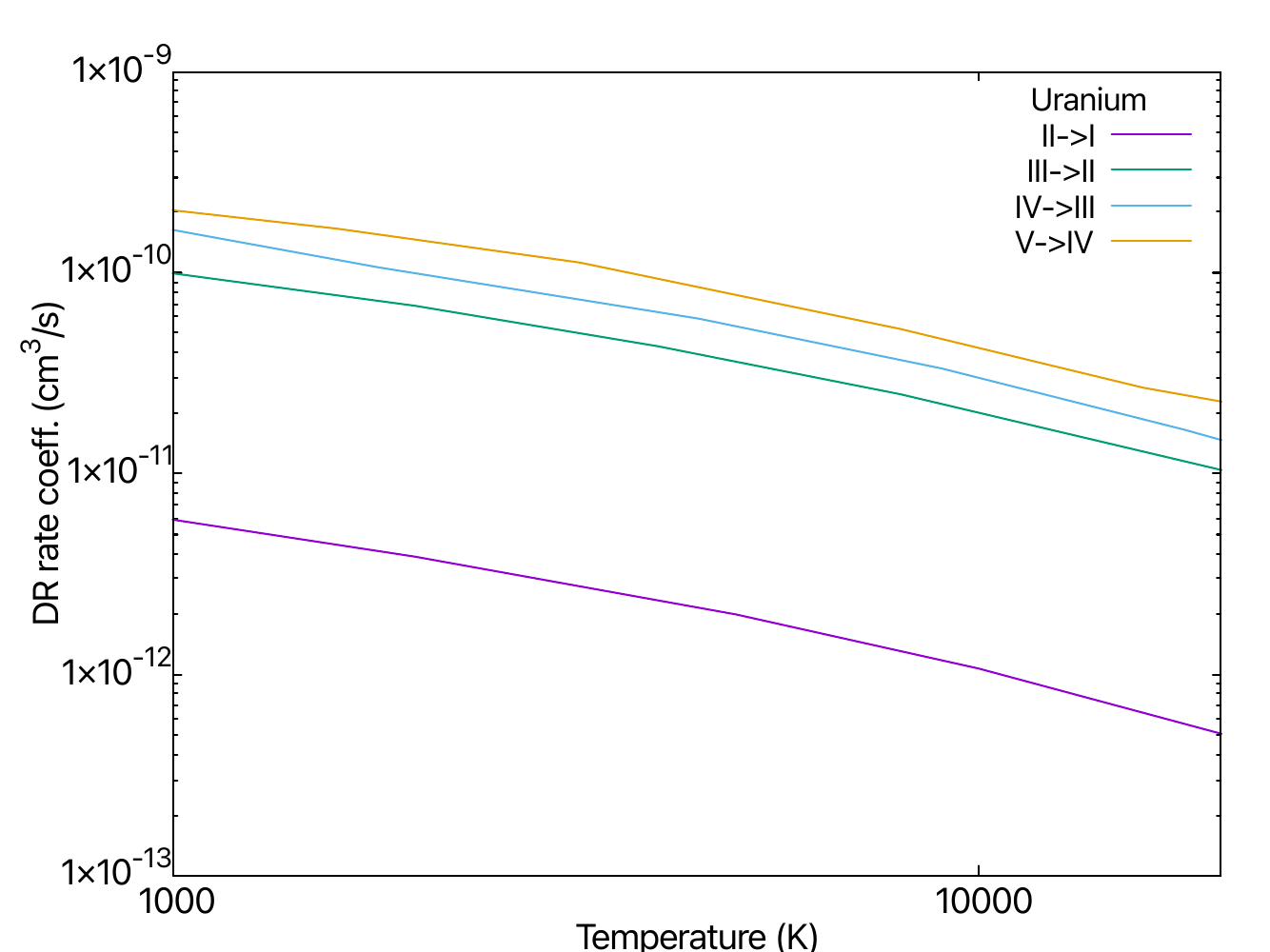}
    \caption{DR rate coefficients of Uranium ions II - V.
    The DR rate coefficients include observed energies from \citet{BlaiseWyart}.}
    \label{fig:UDRAll}
\end{figure}
An actinide DR rate coefficient enables models to incorporate this value to predict an approximate kilonova spectra which will be more accurate than those using the canonical values.
Hence we present DR rate coefficients for ion stages II-V of uranium in figure \ref{fig:UDRAll}.
These rate coefficients were produced in the same way, optimising on the $5f$-shell and applying a final shift in the post-processing stage.

The knowledge gained from initially studying the Nd III f-shell optimised structure and recombination rates was utilized to investigate U II-V.
The atomic structure was the most crucial aspect to consider to ensure resonances were at appropriate positions.
Since there are no previous results to compare with, the recombination rate coefficients presented here were produced from focusing on obtaining an accurate atomic structure and adjusting energies to limited experimental data. 
This is the closest estimation that one can make for the atomic structure and allows an appropriate rate coefficient to be utilized. The amount of experimentally validated levels are limited for these ions but some theoretical predictions by \citet{BlaiseWyart} were used as an estimate.

This simple low temperature optimisation method provided recombination rates that can be applied ot the low temperature applications of kilonova.

\subsection{The effect on Kilonova spectra}
We study the impact of the new recombination rates for Nd III and U ions on kilonova spectra using  the \texttt{SUMO} spectral synthesis code \citep{Jerkstrand2011,Jerkstrand2012}. We compute a  low-$Y_e$ ejecta model, with properties as described in \citet{Pognan2023}, at 30d with the old treatment \citep[constant recombination rates of $10^{-11}$ cm$^3$s$^{-1}$,][]{Pognan22a,Pognan2025} and with the new rates. We solve the rate equations, under the assumption of NLTE and steady state. The ionization equation for ion $x_{j,i}$ (of an element $j$ in ionization state $i$) is given by balancing the ionization ($\Gamma$) and recombination ($\Psi$) flows:
\begin{equation}
\Gamma_{j,i-1} x_{j,i-1} = \Psi_{j,i} x_{j,i}.
\end{equation}
In \texttt{SUMO} ionization occurs via thermal and non-thermal electron collisions and photoionization, and recombination occurs by radiative or resonant dielectronic processes. The ionization and recombination rates generally depend on the ion abundances and the radiation field, and hence, the equations for determining $x_{j,i}$ are non-linear and they are solved by iteration \citep{Jerkstrand11}.

We also replaced the \texttt{FAC} model atoms for U III and U IV in \texttt{SUMO} with the new \texttt{AUTOSTRUCTURE} calculations. U I and U II were not updated as they contribute negligibly to the test model, and also have a very large number of levels (7167 and 3613, respectively) in the \texttt{AUTOSTRUCTURE} model, cumbersome to implement into \texttt{SUMO}. We implemented also the \texttt{AUTOSTRUCTURE}-computed photoionization cross sections for the ground states of Nd II and U I - U III atoms. 

The model has an ejecta mass of 0.05 $M_\odot$, distributed with a $v^{-4}$ power law over the velocity range 0.05-0.3c, in five zones. The composition and radioactive decay power is based on \citet{Wanajo2014}, with thermalization efficiency following the prescriptions of \citet{Kasen2019}.

Even at 30d, the ejecta are still optically thick in the optical due to line opacity. Spectral formation therefore follows a complex series of scattering, absorption, and fluorescence processes. Neodymium is one of the most important agents for this transfer due to its relatively high abundance (2.9\%) and its rich level structure combining to provide a significant fraction of the line opacity. Uranium has a factor $\sim$20 lower abundance (0.15\%) in this model, making it less active in the spectral formation.

The two innermost zones in the model are the most important, accounting for $\sim$95\% of the radioactive decay deposition and, with the highest densities, having the highest line optical depths. The temperatures in these zones are 7300 and 17500 K in the old model, changing only slightly to 8000 K and 18500 K in the new model. This small change is in line with that Nd and U are only two of a large number of elements contributing to the cooling. For similar reasons do the electron number densities change only moderately.

\begin{figure*}
\includegraphics[width=0.49\linewidth]{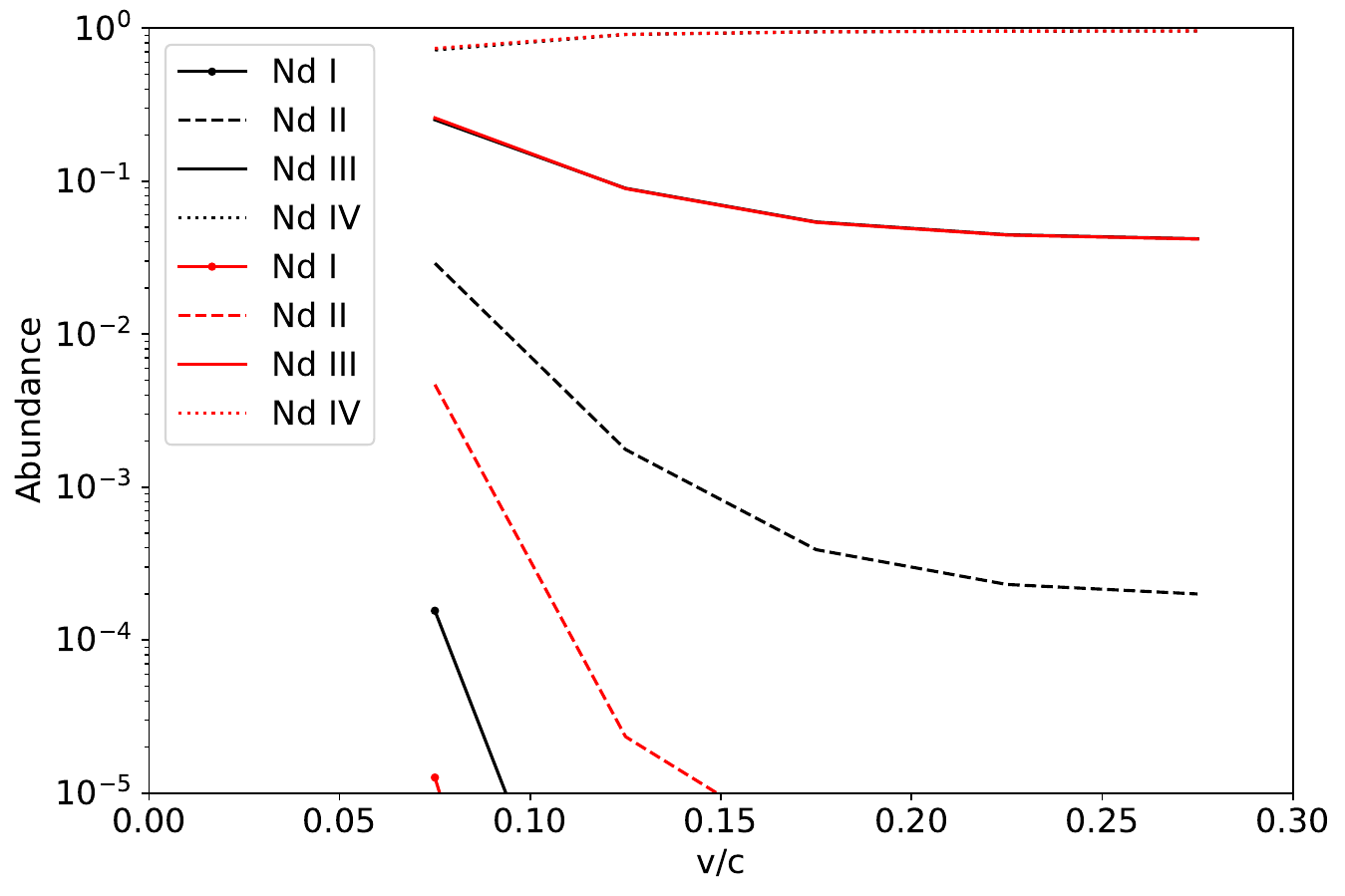}
\includegraphics[width=0.49\linewidth]{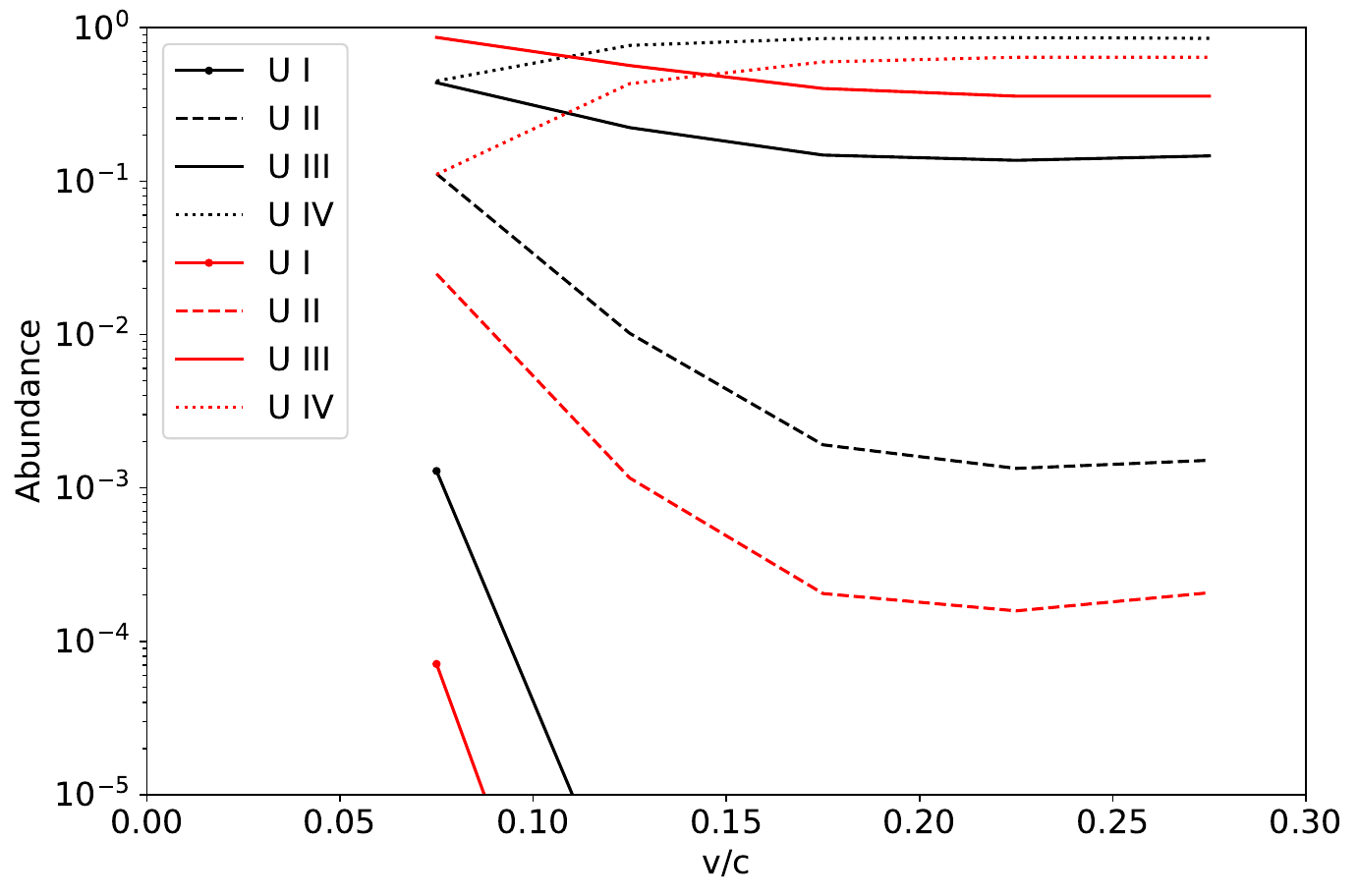}
\caption{Ionization structure for Nd (left) and U (right) in the old model (black) and the new model (red). Neutrals are plotted solid with dots, singly ionized dashed, doubly ionized solid, and triply ionized dotted.}
\label{fig:ionstructure}
\end{figure*}

The ionization solutions for neodymium and uranium change more significantly (Figure. \ref{fig:ionstructure}). In the innermost zone, the old model has Nd I-II-III-IV relative abundances of 
$1.6\times 10^{-4}-0.029-0.25-0.72$, 
whereas the new model has 
$1.3\times 10^{-5}-4.7 \times 10^{-3}-0.26-0.74$.
For U I-II-III-IV, the old model has relative abundances 
$1.3 \times 10^{-3}-0.11-0.44-0.45$,
whereas the new one has 
$7.1\times 10^{-5}-0.025-0.86-0.11$.

\begin{figure*}[ht]
    \centering
    \includegraphics[width=0.8\linewidth]{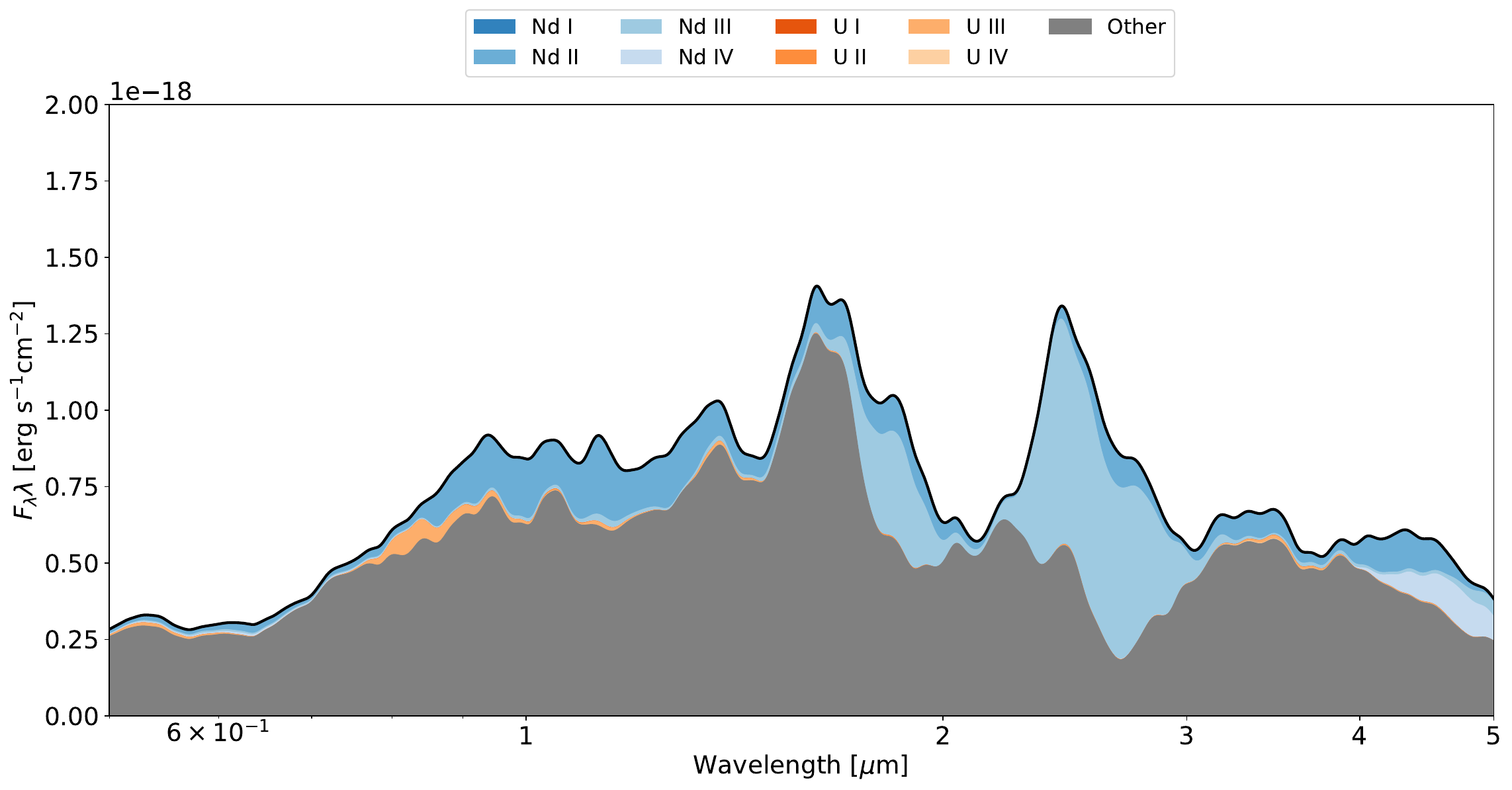}\\
    \includegraphics[width=0.8\linewidth]{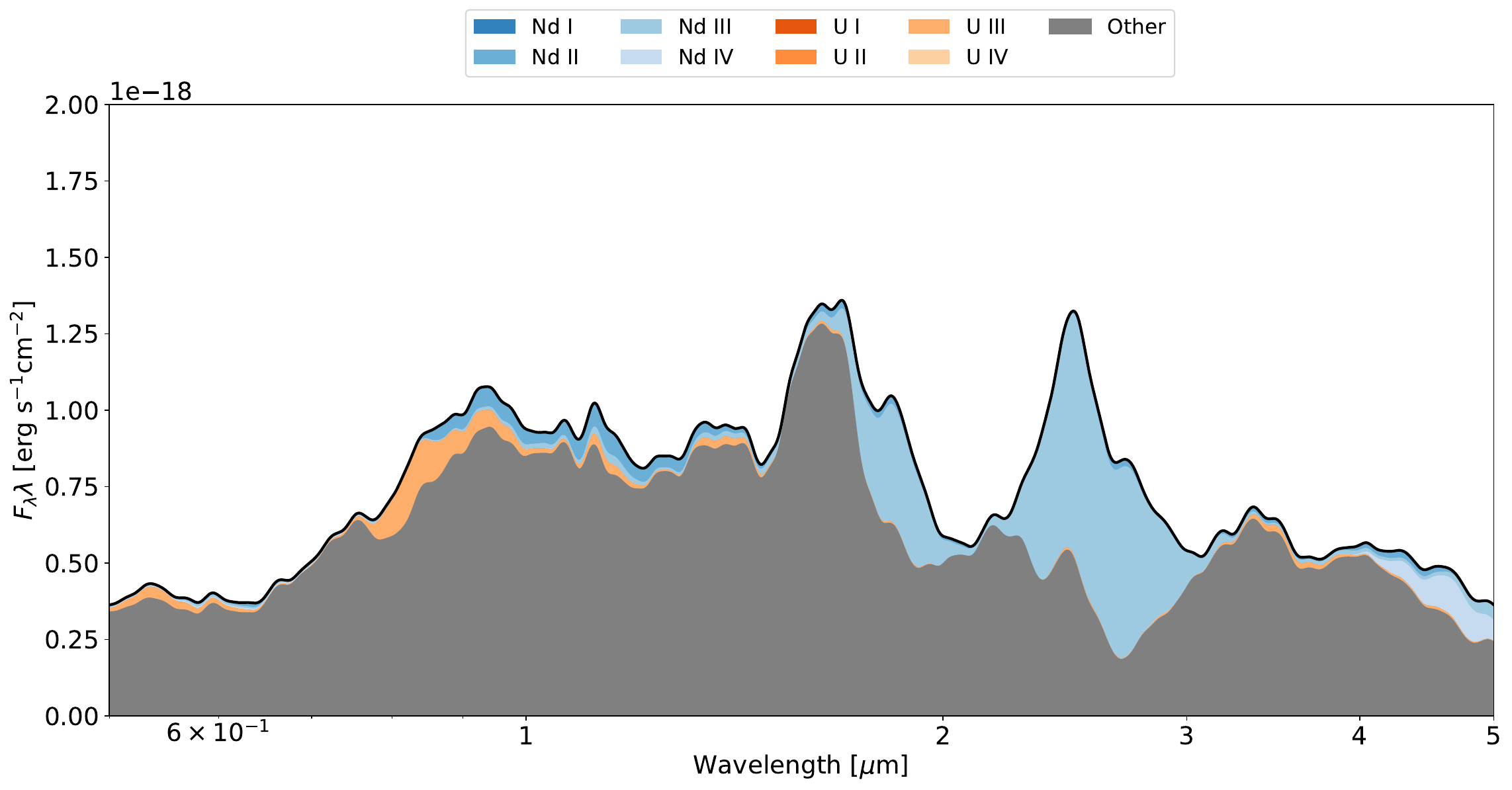}
    \caption{\texttt{SUMO} spectral model at 30d using the old recombination rates (top) and the new ones (bottom). Contributions by Nd and U ions are marked.}
    \label{fig:comparison}
\end{figure*}

Figure \ref{fig:comparison} shows the model with the old rates compared to the new rates. The contributions by various neodymium and uranium ions are marked. In both the old and new models, Nd II and Nd III are active at moderate levels; with Nd II providing most of its flux between $0.8-1.8$ $\mu$m, and Nd III between $1.8-5$ $\mu$m.
The Nd II flux goes down significantly with the new rates, as the abundance is suppressed by a factor of about 10. The Nd III flux is less affected, as its abundance changes only from 0.25 to 0.26 (in the innermost zone). 

Uranium does not have any strong signatures in either of the models, due to its quite low abundance. Other kilonova ejecta models have higher actinide abundances and could potentially give a stronger signal \citep{Pognan2025}. Most of the output is by U III, whose abundance changes from 44\% to 86\% (in the innermost zone) between the models. The relatively modest changes are due to that various U recombination rates are relatively close to the canonical $10^{-11}$ cm$^3$s$^{-1}$ at the particular temperatures in the model ($\gtrsim 10,000$ K). Consequently the signature does not change much between the models. However, other compositions and/or other epochs could give different temperatures and then the deviations from $10^{-11}$ cm$^3$s$^{-1}$ are significant (Fig. \ref{fig:UDRAll}). We note also that these are the first models with reasonably accurate transition wavelengths for U III, showing that its main signatures are at 8,000-10,000 Å.

\section{Discussion and Conclusion} \label{sec:Discussion}

In both cases of the $2^+$ ions of neodymium and uranium, the ground terms are $^5I_{j}$ with $j=4,5,6,7,8$, which means there is fine structure splitting in the ground term. 
An ion with fine-structure splitting in the ground term results in resonances near threshold due to the outer electron stabilisation.
From the E1 selection rule, the allowed dipole transitions associated with the quintet and doublet terms give strong resonances which are seen in the quintet ground term for both Nd III and U III.
The dipole target DR stabilises predominately through the inner electrons as opposed to the outer electron stabilisation which fine-structure DR demonstrates.

The near threshold resonances dominating the DR rate coefficient enables the precision to be aimed towards the smaller energies within 0.7 of a Rydberg. For Nd III and U III, most resonances were within 0.2 Rydberg that were strong enough to contribute to the total DR rate coefficient.
Resonances were spread across the energy range with strong resonance peaks at higher energy, greater than 1 Rydberg.
The higher energy contributions, $E>1$ Rydberg, attribute little as they are only important for high-temperature cases and are outside the temperature range important for kilonovae, allowing for computations to be smaller and quicker.

The dielectronic recombination rate for an f-shell optimized structure of Nd III $\rightarrow$ Nd II was calculated to be $\alpha_{DR}\sim10^{-11}$\ cm$^3$s$^{-1}$ at $T=4000$ K.
Previous calculation by \cite{Hotokezaka} suggested a DR rate coefficient of $\alpha_{DR}\sim9\times10^{-11}$\ cm$^3$s$^{-1}$, at $T=4000$ K.

The dielectronic recombination rate for U III $\rightarrow$ U II was calculated to be $\alpha_{DR}\sim4\times10^{-11}$cm$^3$s$^{-1}$ at $T=4000$ K. 
Both DR rate coefficients are significantly different to the constant recombination rate ($10^{-11}$\ cm$^3$s$^{-1}$) currently being used for modelling confirming the need for temperature dependent rate coefficients.

As the main motivation for this work, the ionization balance depends on the recombination values suggesting various spectral model possibilities.
The difference between the constant recombination rate and calculated dielectronic recombination rates can have a sufficient impact on the ionization balance and spectral signatures, as investigated here for the case of a low-$Y_e$ model at 30d. The new model predicts specific signatures from U III in the $8,000-10,000$ Å range, and from Nd III in the $1.8-3$ micron range.

Domoto et al. (2024) recently explored actinide signatures in early-time kilonova spectra, modelling the first few days after merger under the assumption of local thermodynamic equilibrium (LTE). They found that triply ionised thorium (Th III) can imprint a broad absorption trough around $\lambda \simeq 18,000,\text{\AA}$ when the lanthanide mass fraction is relatively low ($\lesssim 6 \times 10^{-4}$) and the actinide-to-lanthanide ratio exceeds that of the solar $r$-process. Th III is particularly promising because its dense set of low-lying energy levels and large oscillator strengths produce appreciable Sobolev optical depths. The study neglected non-LTE (NLTE) effects, however, leaving open the question of how NLTE physics might affect these spectral features. 

Throughout this investigation, it was clear that the atomic structure of the ion played a large role in determining the recombination rates.
Very small adjustments to the structure caused the DR rate coefficients to vary in values which can create uncertainty between different calculations using the same codes but even more so, different codes calculating recombination rates.
For the Nd III case, different approaches to obtaining the atomic structure and data resulted in a large spread of possibilities for the recombination rate coefficients.
It is clear that the sensitivity with the f-shell has a great impact on the structure and processes.
With differences present between various atomic codes, the recombination rates may differ, but the main difference arises due to the theoretical choice of atomic structure.
Furthermore, the sensitivity to the low-lying energy levels can have significant impact on the rate coefficient.
We have demonstrated recombination rates calculated using \texttt{AUTOSTRUCTURE} \citep{AS} to benchmark results for some ions.
These rate coefficients were optimised for the low-temperature application to kilonovae where the method of structure optimisation was focused to apply to these conditions and only consider the low-lying energies.
Where one would consider higher temperatures, further optimisation beyond the $f$-shell is required and a difference in the high temperature DR rate coefficient would be expected.
Further improvements of the data are expected to have miniscule changes to the final recombination rate but such improvements are possible with the availability of experimental data.
Further investigation into other ion stages of Nd is required due to the sensitivity found in adjustments of the Nd structure. 

It will be beneficial to extend these calculations to other heavy elements that are expected to be present at later times post-merger.
Recombination rate computation should be extended along the lanthanide and actinide group to ensure availability for heavy element recombination with detailed optimization for high temperature features.
Optimization of the f-shell provided most of the structure for the low-temperature applications and so can be a useful method when considering computationally demanding complex species.
Not only will a complete set of atomic data be useful to model spectra with such species, but calculating accurate atomic data to produce recombination rates will improve the accuracy of the spectral models which use them and make element identification much easier from future observations.

\section*{Acknowledgements}
NF acknowledges the Science and Technology Facilities Council (STFC) part of the UK Research and Innovation body for their support via studentship, grant number 2931114. AJ acknowledge support from European Research Council
(ERC) under the European Union’s Horizon 2020 Research and Innovation Program (ERC Starting Grant 803189). The \texttt{SUMO} computations were enabled by resources provided by the National
Academic Infrastructure for Supercomputing in Sweden (NAISS), and the Swedish National Infrastructure for Computing (SNIC), at the Parallelldatorcentrum (PDC) Center for High Performance Computing, Royal Institute of Technology (KTH), partially funded by the Swedish Research Council through grant agreements no. 2022-06725 and no. 2018-05973. SB acknowledges the support from the European Union’s Horizon Europe research and innovation program under the Marie Skłodowska-Curie grant agreement No. 101274945. 

\bibliography{ref}{}

@ARTICLE{Jerkstrand2025,
       author = {{Jerkstrand}, Anders},
        title = "{Spectral synthesis techniques for supernovae and kilonovae}",
      journal = {Living Reviews in Computational Astrophysics},
         year = 2025,
        month = aug,
       volume = {11},
       number = {1},
          eid = {1},
        pages = {1},
          doi = {10.1007/s41115-025-00022-2},
archivePrefix = {arXiv},
       eprint = {2511.17084},
 primaryClass = {astro-ph.HE},
       adsurl = {https://ui.adsabs.harvard.edu/abs/2025LRCA...11....1J}
}

@ARTICLE{Sterling2011-Kr,
       author = {{Sterling}, N.~C.},
        title = "{Atomic data for neutron-capture elements. II. Photoionization and recombination properties of low-charge krypton ions}",
      journal = {\aap},
         year = 2011,
        month = sep,
       volume = {533},
          eid = {A62},
        pages = {A62},
          doi = {10.1051/0004-6361/201117471},
archivePrefix = {arXiv},
       eprint = {1107.3843},
 primaryClass = {astro-ph.SR},
       adsurl = {https://ui.adsabs.harvard.edu/abs/2011A&A...533A..62S}
}

@ARTICLE{Sterling2011-Se,
       author = {{Sterling}, N.~C. and {Witthoeft}, M.~C.},
        title = "{Atomic data for neutron-capture elements. I. Photoionization and recombination properties of low-charge selenium ions}",
      journal = {\aap},
         year = 2011,
        month = may,
       volume = {529},
          eid = {A147},
        pages = {A147},
          doi = {10.1051/0004-6361/201116718},
archivePrefix = {arXiv},
       eprint = {1102.5341},
 primaryClass = {astro-ph.SR},
       adsurl = {https://ui.adsabs.harvard.edu/abs/2011A&A...529A.147S}
}

@ARTICLE{Wollaeger2018,
       author = {{Wollaeger}, Ryan T. and {Korobkin}, Oleg and {Fontes}, Christopher J. and {Rosswog}, Stephan K. and {Even}, Wesley P. and {Fryer}, Christopher L. and {Sollerman}, Jesper and {Hungerford}, Aimee L. and {van Rossum}, Daniel R. and {Wollaber}, Allan B.},
        title = "{Impact of ejecta morphology and composition on the electromagnetic signatures of neutron star mergers}",
      journal = {\mnras},
         year = 2018,
        month = aug,
       volume = {478},
       number = {3},
        pages = {3298-3334},
          doi = {10.1093/mnras/sty1018},
archivePrefix = {arXiv},
       eprint = {1705.07084},
 primaryClass = {astro-ph.HE},
       adsurl = {https://ui.adsabs.harvard.edu/abs/2018MNRAS.478.3298W}
}

@ARTICLE{Tanaka2013,
       author = {{Tanaka}, Masaomi and {Hotokezaka}, Kenta},
        title = "{Radiative Transfer Simulations of Neutron Star Merger Ejecta}",
      journal = {\apj},
         year = 2013,
        month = oct,
       volume = {775},
       number = {2},
          eid = {113},
        pages = {113},
          doi = {10.1088/0004-637X/775/2/113},
archivePrefix = {arXiv},
       eprint = {1306.3742},
 primaryClass = {astro-ph.HE},
       adsurl = {https://ui.adsabs.harvard.edu/abs/2013ApJ...775..113T}
}

@article{DKNB, title={OPACITIES AND SPECTRA OF THE r-PROCESS EJECTA FROM NEUTRON STAR MERGERS}, volume={774}, ISSN={0004-637X}, DOI={10.1088/0004-637X/774/1/25}, number={1}, journal={The Astrophysical Journal}, publisher={The American Astronomical Society}, author={Kasen, Daniel and Badnell, N. R. and Barnes, Jennifer}, year={2013}, month=aug, pages={25}, language={en} }

@ARTICLE{Jerkstrand2012,
       author = {{Jerkstrand}, A. and {Fransson}, C. and {Maguire}, K. and {Smartt}, S. and {Ergon}, M. and {Spyromilio}, J.},
        title = "{The progenitor mass of the Type IIP supernova SN 2004et from late-time spectral modeling}",
      journal = {\aap},
         year = 2012,
        month = oct,
       volume = {546},
          eid = {A28},
        pages = {A28},
          doi = {10.1051/0004-6361/201219528},
archivePrefix = {arXiv},
       eprint = {1208.2183},
 primaryClass = {astro-ph.HE},
       adsurl = {https://ui.adsabs.harvard.edu/abs/2012A&A...546A..28J}
}

@ARTICLE{Jerkstrand2011,
       author = {{Jerkstrand}, A. and {Fransson}, C. and {Kozma}, C.},
        title = "{The $^{44}$Ti-powered spectrum of SN 1987A}",
      journal = {\aap},
         year = 2011,
        month = jun,
       volume = {530},
          eid = {A45},
        pages = {A45},
          doi = {10.1051/0004-6361/201015937},
archivePrefix = {arXiv},
       eprint = {1103.3653},
 primaryClass = {astro-ph.HE},
       adsurl = {https://ui.adsabs.harvard.edu/abs/2011A&A...530A..45J}
}

@ARTICLE{Kasen2019,
       author = {{Kasen}, Daniel and {Barnes}, Jennifer},
        title = "{Radioactive Heating and Late Time Kilonova Light Curves}",
      journal = {\apj},
         year = 2019,
        month = may,
       volume = {876},
       number = {2},
          eid = {128},
        pages = {128},
          doi = {10.3847/1538-4357/ab06c2},
archivePrefix = {arXiv},
       eprint = {1807.03319},
 primaryClass = {astro-ph.HE},
       adsurl = {https://ui.adsabs.harvard.edu/abs/2019ApJ...876..128K}
}

@ARTICLE{Wanajo2014,
       author = {{Wanajo}, Shinya and {Sekiguchi}, Yuichiro and {Nishimura}, Nobuya and {Kiuchi}, Kenta and {Kyutoku}, Koutarou and {Shibata}, Masaru},
        title = "{Production of All the r-process Nuclides in the Dynamical Ejecta of Neutron Star Mergers}",
      journal = {\apjl},
         year = 2014,
        month = jul,
       volume = {789},
       number = {2},
          eid = {L39},
        pages = {L39},
          doi = {10.1088/2041-8205/789/2/L39},
archivePrefix = {arXiv},
       eprint = {1402.7317},
 primaryClass = {astro-ph.SR},
       adsurl = {https://ui.adsabs.harvard.edu/abs/2014ApJ...789L..39W}
}

@ARTICLE{Jerkstrand11,
       author = {{Jerkstrand}, A. and {Fransson}, C. and {Kozma}, C.},
        title = "{The $^{44}$Ti-powered spectrum of SN 1987A}",
      journal = {\aap},
         year = 2011,
        month = jun,
       volume = {530},
          eid = {A45},
        pages = {A45},
          doi = {10.1051/0004-6361/201015937},
archivePrefix = {arXiv},
       eprint = {1103.3653},
 primaryClass = {astro-ph.HE},
       adsurl = {https://ui.adsabs.harvard.edu/abs/2011A&A...530A..45J}
}

@ARTICLE{banerjee25,
       author = {{Banerjee}, Smaranika and {Jerkstrand}, Anders and {Badnell}, Nigel and {Pognan}, Quentin and {Ferguson}, Niamh and {Grumor}, Jon},
        title = "{Nebular spectra of kilonovae with detailed recombination rates -- I. Light r-process composition}",
      journal = {arXiv e-prints},
         year = 2025,
        month = jan,
          eid = {arXiv:2501.18345},
        pages = {arXiv:2501.18345},
          doi = {10.48550/arXiv.2501.18345},
archivePrefix = {arXiv},
       eprint = {2501.18345},
 primaryClass = {astro-ph.HE},
       adsurl = {https://ui.adsabs.harvard.edu/abs/2025arXiv250118345B}
}

@ARTICLE{Pognan22a,
       author = {{Pognan}, Quentin and {Jerkstrand}, Anders and {Grumer}, Jon},
        title = "{On the validity of steady-state for nebular phase kilonovae}",
      journal = {\mnras},
         year = 2022,
        month = mar,
       volume = {510},
       number = {3},
        pages = {3806-3837},
          doi = {10.1093/mnras/stab3674},
archivePrefix = {arXiv},
       eprint = {2112.07484},
 primaryClass = {astro-ph.HE},
       adsurl = {https://ui.adsabs.harvard.edu/abs/2022MNRAS.510.3806P}
}

@ARTICLE{Pognan2023,
       author = {{Pognan}, Quentin and {Grumer}, Jon and {Jerkstrand}, Anders and {Wanajo}, Shinya},
        title = "{NLTE spectra of kilonovae}",
      journal = {\mnras},
         year = 2023,
        month = dec,
       volume = {526},
       number = {4},
        pages = {5220-5248},
          doi = {10.1093/mnras/stad3106},
archivePrefix = {arXiv},
       eprint = {2309.01134},
 primaryClass = {astro-ph.HE},
       adsurl = {https://ui.adsabs.harvard.edu/abs/2023MNRAS.526.5220P}
}

@article{DK2017, title={Origin of the heavy elements in binary neutron-star mergers from a gravitational-wave event}, volume={551}, rights={2017 Macmillan Publishers Limited, part of Springer Nature. All rights reserved.}, ISSN={1476-4687}, DOI={10.1038/nature24453}, number={7678}, journal={Nature}, publisher={Nature Publishing Group}, author={Kasen, Daniel and Metzger, Brian and Barnes, Jennifer and Quataert, Eliot and Ramirez-Ruiz, Enrico}, year={2017}, month=nov, pages={80–84}, language={en} }

@article{FontesLa, title={A line-binned treatment of opacities for the spectra and light curves from neutron star mergers}, volume={493}, ISSN={0035-8711}, DOI={10.1093/mnras/staa485}, number={3}, journal={Monthly Notices of the Royal Astronomical Society}, author={Fontes, C J and Fryer, C L and Hungerford, A L and Wollaeger, R T and Korobkin, O}, year={2020}, month=apr, pages={4143–4171} }

@article{FontesAc, title={Actinide opacities for modelling the spectra and light curves of kilonovae}, volume={519}, ISSN={0035-8711}, DOI={10.1093/mnras/stac2792}, number={2}, journal={Monthly Notices of the Royal Astronomical Society}, author={Fontes, C J and Fryer, C L and Wollaeger, R T and Mumpower, M R and Sprouse, T M}, year={2023}, month=feb, pages={2862–2878} }

@article{Hotokezaka, title={Nebular emission from lanthanide-rich ejecta of neutron star merger}, volume={506}, ISSN={0035-8711}, DOI={10.1093/mnras/stab1975}, number={4}, journal={Monthly Notices of the Royal Astronomical Society}, author={Hotokezaka, Kenta and Tanaka, Masaomi and Kato, Daiji and Gaigalas, Gediminas}, year={2021}, month=oct, pages={5863–5877} }

@article{Brewer, title={Energies of the Electronic Configurations of the Singly, Doubly, and Triply Ionized Lanthanides and Actinides}, volume={61}, rights={© 1971 Optical Society of America}, DOI={10.1364/JOSA.61.001666}, number={12}, journal={JOSA}, publisher={Optica Publishing Group}, author={Brewer, Leo}, year={1971}, month=dec, pages={1666–1682}, language={EN} }

@article{PognanNLTE, title={NLTE spectra of kilonovae}, volume={526}, ISSN={0035-8711}, DOI={10.1093/mnras/stad3106}, number={4}, journal={Monthly Notices of the Royal Astronomical Society}, author={Pognan, Quentin and Grumer, Jon and Jerkstrand, Anders and Wanajo, Shinya}, year={2023}, month=dec, pages={5220–5248} }

@article{DR, title={Dielectronic recombination data for dynamic finite-density plasmas - I. Goals and methodology}, volume={406}, rights={© ESO, 2003}, ISSN={0004-6361, 1432-0746}, DOI={10.1051/0004-6361:20030816}, number={33}, journal={Astronomy \& Astrophysics}, publisher={EDP Sciences}, author={Badnell, N. R. and O’Mullane, M. G. and Summers, H. P. and Altun, Z. and Bautista, M. A. and Colgan, J. and Gorczyca, T. W. and Mitnik, D. M. and Pindzola, M. S. and Zatsarinny, O.}, year={2003}, month=aug, pages={1151–1165}, language={en} }

@article{RR, title={Radiative Recombination Data for Modeling Dynamic Finite-Density Plasmas}, volume={167}, ISSN={0067-0049}, DOI={10.1086/508465},  number={2}, journal={The Astrophysical Journal Supplement Series}, publisher={IOP Publishing}, author={Badnell, N. R.}, year={2006}, month=dec, pages={334}, language={en} }

@article{FeDR, title={Dielectronic recombination of Fe13+: benchmarking the M-shell}, volume={39}, ISSN={0953-4075}, DOI={10.1088/0953-4075/39/23/003}, abstractNote={We have carried-out a series of multi-configuration Breit–Pauli AUTOSTRUCTURE calculations for the dielectronic recombination of Fe13+. We present a detailed comparison of the results with the high-energy-resolution measurements reported recently from the Heidelberg heavy-ion Test Storage Ring by Schmidt et al. Many Rydberg series contribute significantly from this initial 3s23p M-shell ion, resulting in a complex recombination ‘spectrum’. While there is much close agreement between theory and experiment, differences of typically 50% in the summed resonance strengths over 0.1–10 eV result in the experimentally based total Maxwellian recombination rate coefficient being a factor of 1.52–1.38 larger than theory over 104–105 K, which is a typical temperature range of peak abundance for Fe13+ in a photoionized plasma. Nevertheless, the present theoretical recombination rate coefficient is an order of magnitude larger than that used by modellers to date. This may help explain the discrepancy between the iron M-shell ionization balance predicted by photoionization modelling codes, such as ION and CLOUDY, and that deduced from the iron M-shell unresolved-transition-array absorption feature observed in the x-ray spectrum of many active galactic nuclei. Similar data are required for Fe8+ through Fe12+ in order to remove the question mark hanging over the atomic data though.}, number={23}, journal={Journal of Physics B: Atomic, Molecular and Optical Physics}, author={Badnell, N. R.}, year={2006}, month=nov, pages={4825}, language={en} }

@book{BlaiseWyart, address={France}, series={Constantes selectionnees niveaux d’energie et spectres atomiques des actinides}, title={Selected constants energy levels and atomic spectra of actinides}, ISBN={978-2-9506414-0-3}, abstractNote={In this book is given a compilation of experimental data on energy levels and atomic
spectra of neutral and ionized actinides: actinium, thorium, protactinium, uranium,
neptunium, plutonium, americium, curium, berkelium, californium and einsteinium}, note={INIS Reference Number: 24031406}, publisher={Centre National de la Recherche Scientifique}, author={Blaise, J. and Wyart, J.F.}, year={1992}, collection={Constantes selectionnees niveaux d’energie et spectres atomiques des actinides} }

@book{NISTNd, address={Gaithersburg, MD}, title={Atomic energy levels - the rare-earth elements: the spectra of lanthanum, cerium, praseodymium, neodymium, promethium, samarium, europium, gadolinium, terbium, dysprosium, holmium, erbium, thulium, ytterbium, and lutetitum}, url={https://nvlpubs.nist.gov/nistpubs/Legacy/NSRDS/nbsnsrds60.pdf}, DOI={10.6028/NBS.NSRDS.60}, institution={National Bureau of Standards}, author={Martin, W C and Zalubas, Romuald and Hagan, Lucy}, year={1978}, language={en} }

@article{AS, title={A Breit–Pauli distorted wave implementation for autostructure}, volume={182}, ISSN={0010-4655}, DOI={10.1016/j.cpc.2011.03.023}, number={7}, journal={Computer Physics Communications}, author={Badnell, N. R.}, year={2011}, month=jul, pages={1528–1535} }

@article{Wanajo_2018, title={Physical Conditions for the r-process. I. Radioactive Energy Sources of Kilonovae}, volume={868}, ISSN={0004-637X}, DOI={10.3847/1538-4357/aae0f2}, abstractNote={Radioactive energies from unstable nuclei made in the ejecta of neutron star mergers play principal roles in powering kilonovae. In previous studies, power-law-type heating rates (e.g., ) have frequently been used, which may be inadequate if the ejecta are dominated by nuclei other than the A ∼ 130 region. We consider, therefore, two reference abundance distributions that match the r-process residuals to the solar abundances for A ≥ 69 (light trans-iron plus r-process elements) and A ≥ 90 (r-process elements). Nucleosynthetic abundances are obtained by using free-expansion models with three parameters: expansion velocity, entropy, and electron fraction. Radioactive energies are calculated as an ensemble of weighted free-expansion models that reproduce the reference abundance patterns. The results are compared with the bolometric luminosity (> a few days since merger) of the kilonova associated with GW170817. We find that the former case (fitted for A ≥ 69) with an ejecta mass 0.06 M⊙ reproduces the light curve remarkably well, including its steepening at ≳7 days, in which the mass of r-process elements is ≈0.01 M⊙. Two β-decay chains are identified: 66Ni 66Cu 66Zn and 72Zn 72Ga 72Ge with similar halflives of parent isotopes (≈2 days), which leads to an exponential-like evolution of heating rates during 1–15 days. The light curve at late times (>40 days) is consistent with additional contributions from the spontaneous fission of 254Cf and a few Fm isotopes. If this is the case, the GW170817 event is best explained by the production of both light trans-iron and r-process elements that originate from dynamical ejecta and subsequent disk outflows from the neutron star merger.}, number={1}, journal={The Astrophysical Journal}, publisher={The American Astronomical Society}, author={Wanajo, Shinya}, year={2018}, month=nov, pages={65}, language={en} }

@article{Pognan2025, title={Actinide signatures in low electron fraction kilonova ejecta}, volume={536}, ISSN={0035-8711}, DOI={10.1093/mnras/stae2778}, abstractNote={Neutron star (NS) mergers are known to produce heavy elements through rapid neutron capture (r-process) nucleosynthesis. Actinides are expected to be created solely by the r-process in the most neutron-rich environments. Confirming if NS mergers provide the requisite conditions for actinide creation is therefore central to determining their origin in the Universe. Actinide signatures in kilonova (KN) spectra may yield an answer, provided adequate models are available in order to interpret observational data. In this study, we investigate actinide signatures in neutron-rich merger ejecta. We use three ejecta models with different compositions and radioactive power, generated by nucleosynthesis calculations using the same initial electron fraction ($Y_e = 0.15$) but with different nuclear physics inputs and thermodynamic expansion history. These are evolved from 10 to 100 d after merger using the sumo non-local thermodynamic equilibrium (NLTE) radiative transfer code. We highlight how uncertainties in nuclear properties, as well as choices in thermodynamic trajectory, may yield entirely different outputs for equal values of $Y_e$. We consider an actinide-free model and two actinide-rich models, and find that the emergent spectra and light-curve evolution are significantly different depending on the amount of actinides present, and the overall decay properties of the models. We also present potential key actinide spectral signatures, of which doubly ionized $_{89}$Ac and $_{90}$Th may be particularly interesting as spectral indicators of actinide presence in KN ejecta.}, number={3}, journal={Monthly Notices of the Royal Astronomical Society}, author={Pognan, Quentin and Wu, Meng-Ru and Martínez-Pinedo, Gabriel and da Silva, Ricardo Ferreira and Jerkstrand, Anders and Grumer, Jon and Flörs, Andreas}, year={2025}, month=jan, pages={2973–2992} }

@article{Mulholland_Ramsbottom_Ballance_Sneppen_Sim_2026, title={Electron impact excitation of Te iv and v and level resolved R-matrix photoionization of Te i–iv with application to modelling of AT2017gfo}, volume={546}, ISSN={0035-8711}, DOI={10.1093/mnras/stag237}, abstractNote={Spectral modelling of kilonovae (KNe) requires large amounts of collisional excitation and photoionization atomic data for lowly ionized (neutral, singly, and doubly ionized) species of heavy elements. Much of the data currently used is calculated using approximate hydrogenic results or adopts semi-empirical formulae. We present atomic data for ions of tellurium (Te) computed using the well-known R-matrix method. Results will also be presented for radiative and thermal collisions of Te iv and v, for which the required atomic data are also typically limited in the literature. The Multi-Configuration-Dirac–Hartree–Fock method is used to produce model atomic structures and radiative rates. These model structures are then used to calculate electron-impact-excitation and photoionization cross-sections. The resulting excitation and radiative rates are further used in a collisional radiative model to produce synthetic spectra, which are compared with observations. We also investigate the possibility of Te iv contributing to the 1.08 $mu$m emission feature in the mid-epochs of AT2017gfo alongside the established P-Cygni feature of Sr ii.}, number={4}, journal={Monthly Notices of the Royal Astronomical Society}, author={Mulholland, Leo P and Ramsbottom, Catherine A and Ballance, Connor P and Sneppen, Albert and Sim, Stuart A}, year={2026}, month=mar, pages={stag237} }

@article{McCann_Ballance_McNeill_Sim_Ramsbottom_2025, title={Electron-impact excitation of zirconium i–iii in support of neutron star merger diagnostics}, volume={540}, ISSN={0035-8711}, DOI={10.1093/mnras/staf866}, abstractNote={Recent observations and analyses of kilonova spectra as a result of neutron star mergers require accurate and complete atomic structure and collisional data for interpretation. Ideally, the atomic data sets for elements predicted to be abundant in the ejecta should be experimentally calibrated. For near-neutral ion stages of zirconium in particular, the A-values and the associated excitation/de-excitation rates are required from collision calculations built upon accurate structure models. The atomic orbitals required to perform the structure calculations may be calculated using a Multi-Configuration-Dirac-Fock (MCDF) approximation implemented within the General Relativistic Atomic Structure Package (grasp0). Optimized sets of relativistic atomic orbitals are then imported into electron-impact excitation collision calculations. A relativistic R-matrix formulation within the Dirac Atomic R-matrix Code (darc) is employed to compute collision strengths, which are subsequently Maxwellian convolved to produce excitation/de-excitation rates for a wide range of electron temperatures. These atomic data sets subsequently provide the foundations for non-local thermodynamic equilibrium collisional-radiative models. In this work, all these computations have been carried out for the first three ion stages of zirconium (Zr i–iii) with the data further interfaced with collisional-radiative and radiative transfer codes to produce synthetic spectra that can be compared with observation.}, number={4}, journal={Monthly Notices of the Royal Astronomical Society}, author={McCann, M and Ballance, C P and McNeill, F and Sim, S A and Ramsbottom, C A}, year={2025}, pages={2923–2936} }

@article{Jerkstrand2026, title={Infrared spectral signatures of light r -process elements in kilonovae}, volume={548}, rights={https://creativecommons.org/licenses/by/4.0/}, ISSN={0035-8711, 1365-2966}, DOI={10.1093/mnras/stag733}, abstractNote={ABSTRACT
            A central question regarding neutron star (NS) mergers is whether they are able to produce all the r-process elements, from first to third peak. We here study theoretical infrared signatures of first-peak elements with spectral synthesis modelling. By combining state-of-the-art non-local thermodynamic equilibrium physics with new radiative and collisional data for these elements, we identify several promising diagnostic lines from Ge, As, Se, Br, Kr, and Zr. The models give self-consistent line luminosities and indicate specific features that probe emission volumes at early phases ($sim$10 d), the product of ion mass and electron density in late phases ($gtrsim$75 d), and in some cases direct ionic masses at intermediate phases. Emission by [Se i] 5.03 $mu$m + [Se iii] 4.55 $mu$m is the only candidate from the first r-process peak that could explain the Spitzer photometry of AT2017gfo. However, the models show consistently that with a Kr/Te and Se/Te ratio following the solar r-process pattern, Kr + Se emission is dominant over Te for the feature at 2.1 $mu$m observed in both AT2017gfo and AT2023vfi. The somewhat better line profile fit with [Te iii] may suggest that both AT2017gfo and AT2023vfi had a strongly subsolar production of the light r-process elements. An alternative scenario could be that Kr + Se in an asymmetric morphological distribution generates the feature. Further James Webb Space Telescope spectral observations hold promise to determine the light r-process production of kilonovae, and in particular whether the light elements are made in a slow disc outflow or in a fast proto-NS wind. We identify specific needs for further atomic data for $Z=31-40$ elements.}, number={4}, journal={Monthly Notices of the Royal Astronomical Society}, author={Jerkstrand, Anders and Pognan, Quentin and Banerjee, Smaranika and Sterling, Nicholas C and Grumer, Jon and Ferguson, Niamh and Butler, Keith and Gillanders, James and Smartt, Stephen and Kawaguchi, Kyohei and Vilagos, Blanka}, year={2026}, month=may, pages={stag733}, language={en} }

@article{Carvajal_Gallego_Berengut_Palmeri_Quinet_2021, title={Large-scale atomic data calculations in Ce V – X ions for application to early kilonova emission from neutron star mergers}, volume={509}, rights={https://academic.oup.com/journals/pages/open_access/funder_policies/chorus/standard_publication_model}, ISSN={0035-8711, 1365-2966}, DOI={10.1093/mnras/stab3423}, number={4}, journal={Monthly Notices of the Royal Astronomical Society}, author={Carvajal Gallego, H and Berengut, J C and Palmeri, P and Quinet, P}, year={2021}, month=dec, pages={6138–6154}, language={en} }

@article{Gillanders_Smartt_2025, title={Analysis of the JWST spectra of the kilonova AT 2023vfi accompanying GRB 230307A}, volume={538}, ISSN={0035-8711}, DOI={10.1093/mnras/staf287}, abstractNote={Kilonovae are key to advancing our understanding of r-process nucleosynthesis. To date, only two kilonovae have been spectroscopically observed, AT 2017gfo and AT 2023vfi. Here, we present an analysis of the JWST spectra obtained +29 and +61 d post-merger for AT 2023vfi (the kilonova associated with GRB 230307A). After re-reducing and photometrically flux-calibrating the data, we empirically model the observed X-ray to mid-infrared continua with a power law and a blackbody, to replicate the non-thermal afterglow and apparent thermal continuum $gtrsim 2$ $mu$m. We fit Gaussians to the apparent emission features, obtaining line centroids of $20218_{-38}^{+37}$, $21874 pm 89$, and $44168_{-152}^{+153}$ Å, and velocity widths spanning $0.057 - 0.110$ c. These line centroid constraints facilitated a detailed forbidden line identification search, from which we shortlist a number of r-process species spanning all three r-process peaks. We rule out Ba ii and Ra ii as candidates and propose Te i–iii, Er i–iii, and W iii as the most promising ions for further investigation, as they plausibly produce multiple emission features from one (W iii) or multiple (Te i–iii, Er i–iii) ion stages. We compare to the spectra of AT 2017gfo, which also exhibit prominent emission at $sim 2.1$ $mu$m, and conclude that [Te iii] $lambda 21050$ remains the most plausible cause of the observed $sim 2.1$ $mu$m emission in both kilonovae. However, the observed line centroids are not consistent between both objects, and they are significantly offset from [Te iii] $lambda 21050$. The next strongest [Te iii] transition at 29 290 Å is not observed, and we quantify its detectability. Further study is required, with particular emphasis on expanding the available atomic data to enable quantitative non-LTE spectral modelling.}, number={3}, journal={Monthly Notices of the Royal Astronomical Society}, author={Gillanders, J H and Smartt, S J}, year={2025}, month=apr, pages={1663–1689} }

@article{Deprince_Wagle_Nasr_Gallego_Godefroid_Goriely_Just_Palmeri_Quinet_Eck_2025, title={Kilonova ejecta opacity inferred from new large-scale HFR atomic calculations in all elements between Ca (Z = 20) and Lr (Z = 103)}, volume={696}, rights={© The Authors 2025}, ISSN={0004-6361, 1432-0746}, DOI={10.1051/0004-6361/202452967}, abstractNote={<i>Context.<i/> The production of elements heavier than iron in the Universe still remains an unsolved mystery. About half of them are thought to be produced by the astrophysical r-process (rapid neutron-capture process), for which neutron star mergers (NSMs) are among the most promising production sites. In August 2017, gravitational waves generated by a NSM were detected for the first time by the LIGO detectors (event GW170817), and the observation of its electromagnetic counterpart, the kilonova (KN) AT2017gfo, suggested the presence of heavy elements in the KN ejecta. The luminosity and spectra of such KN emission depend significantly on the ejecta opacity. Atomic data and opacities for heavy elements are thus sorely needed to model and interpret KN light curves and spectra.<i>Aims.<i/> The present work focuses on large-scale atomic data and opacity computations for all heavy elements with <i>Z<i/> ≥ 20, with a special effort on lanthanides and actinides, for a grid of typical KN ejecta conditions (temperature, density, and time post-merger) between one day and one week after the merger (corresponding to the local thermodynamical equilibrium photosphere phase of the KN ejecta).<i>Methods.<i/> In order to do so, we used the pseudo-relativistic Hartree–Fock (HFR) method as implemented in Cowan’s codes, in which the choice of the interaction configuration model is of crucial importance.<i>Results.<i/> In this paper, HFR atomic data and opacities for all elements between Ca (<i>Z<i/> = 20) and Lr (<i>Z<i/> = 103) are presented, with a special focus on lanthanides and actinides. In particular, we found increased lanthanide opacities compared to previous works. Besides, we also discuss the contribution of every single element with <i>Z<i/> ≥ 20 to the total KN ejecta opacity for a given NSM model, depending on their Planck mean opacities and elemental abundances. An important result is that lanthanides are found to not be the dominant sources of opacity, at least on average. The impact on KN light curves of considering such atomic-physics-based opacity data instead of typical crude approximation formulae is also evaluated. In addition, the importance of taking the ejecta composition into account directly in the expansion opacity determination (instead of estimating single-element opacities) is highlighted. A database containing all the relevant atomic data and opacity tables has also been created and published online along with this work.}, journal={Astronomy \& Astrophysics}, publisher={EDP Sciences}, author={Deprince, J. and Wagle, G. and Nasr, S. Ben and Gallego, H. Carvajal and Godefroid, M. and Goriely, S. and Just, O. and Palmeri, P. and Quinet, P. and Eck, S. Van}, year={2025}, month=apr, pages={A32}, language={en} }

@article{Burgess_1964, title={Delectronic Recombination and the Temperature of the Solar Corona.}, volume={139}, DOI={10.1086/147813}, journal={The Astrophysical Journal}, author={Burgess, Alan}, year={1964}, month=feb, pages={776–780} }

@article{Banerjee, title={Opacity of the Highly Ionized Lanthanides and the Effect on the Early Kilonova}, volume={934}, ISSN={0004-637X}, DOI={10.3847/1538-4357/ac7565}, number={2}, journal={The Astrophysical Journal}, publisher={The American Astronomical Society}, author={Banerjee, Smaranika and Tanaka, Masaomi and Kato, Daiji and Gaigalas, Gediminas and Kawaguchi, Kyohei and Domoto, Nanae}, year={2022}, month=jul, pages={117}, language={en} }

@article{Experimental, title={Determination of the plasma recombination rate coefficient}, volume={409}, rights={© ESO, 2003}, ISSN={0004-6361, 1432-0746}, DOI={10.1051/0004-6361:20031100}, abstractNote={The absolute radiative and dielectronic recombination rate coefficients for were measured at the CRYRING storage ring for the center-of-mass energy range 0–43 eV, which covers all core excitations up to the 3p<sub>1/2<sub/> and 3p<sub>3/2<sub/> series limits. These are compared to an AUTOSTRUCTURE intermediate coupling calculation. The absolute rate coefficients are then convoluted with a Maxwellian temperature distribution to obtain the temperature dependent plasma recombination rate coefficient for the temperature range 10<sup>3<sup/>–10<sup>7<sup/> K. The temperature dependent rate coefficient can be useful as a tool in both astrophysical and laboratory plasma diagnostics where the knowledge of ionization balance and level populations are of interest in deciphering the many emission lines observed and for the evaluation of the plasma temperature via emission line ratios.}, number={22}, journal={Astronomy \& Astrophysics}, publisher={EDP Sciences}, author={Fogle, M. and Badnell, N. R. and Eklöw, N. and Mohamed, T. and Schuch, R.}, year={2003}, month=oct, pages={781–786}, language={en} }

@article{Tanaka2020, title={Systematic opacity calculations for kilonovae}, volume={496}, ISSN={0035-8711}, DOI={10.1093/mnras/staa1576}, number={2}, journal={Monthly Notices of the Royal Astronomical Society}, author={Tanaka, Masaomi and Kato, Daiji and Gaigalas, Gediminas and Kawaguchi, Kyohei}, year={2020}, month=aug, pages={1369–1392} }

@article{banerjee_simulations_2020,
    title = {Simulations of {Early} {Kilonova} {Emission} from {Neutron} {Star} {Mergers}},
    volume = {901},
    issn = {0004-637X},
    url = {https://dx.doi.org/10.3847/1538-4357/abae61},
    doi = {10.3847/1538-4357/abae61},
    language = {en},
    number = {1},
    urldate = {2024-10-31},
    journal = {The Astrophysical Journal},
    author = {Banerjee, Smaranika and Tanaka, Masaomi and Kawaguchi, Kyohei and Kato, Daiji and Gaigalas, Gediminas},
    month = sep,
    year = {2020},
    note = {Publisher: The American Astronomical Society},
    pages = {29},
}

@article{banerjee_diversity_2024,
    title = {Diversity of {Early} {Kilonova} with the {Realistic} {Opacities} of {Highly} {Ionized} {Heavy} {Elements}},
    volume = {968},
    issn = {0004-637X},
    url = {https://dx.doi.org/10.3847/1538-4357/ad4029},
    doi = {10.3847/1538-4357/ad4029},
    language = {en},
    number = {2},
    urldate = {2025-06-09},
    journal = {The Astrophysical Journal},
    author = {Banerjee, Smaranika and Tanaka, Masaomi and Kato, Daiji and Gaigalas, Gediminas},
    month = jun,
    year = {2024},
    note = {Publisher: The American Astronomical Society},
    pages = {64},
}

@article{Ding_Ryabtsev_Kononov_Ryabchikova_Clear_Concepcion_Pickering_2024, title={Spectrum and energy levels of the low-lying configurations of Nd III}, volume={684}, rights={https://creativecommons.org/licenses/by/4.0}, ISSN={0004-6361, 1432-0746}, DOI={10.1051/0004-6361/202348794}, abstractNote={Methods. The emission spectra of neodymium (Nd, Z = 60) were recorded using Penning and hollow cathode discharge lamps in the region 11 500–54 000 cm−1 (8695–1852 Å) by Fourier transform spectroscopy at resolving powers up to 106. Wavenumber measurements were accurate to a few 10−3 cm−1. Grating spectroscopy of Nd vacuum sliding sparks and stellar spectra were used to aid line and energy level identification. For the analysis, new Nd III atomic structure and transition probability calculations were carried out using the Cowan code parameterised by newly established levels.
Results. The classification of 432 transitions of Nd III from the Penning lamp spectra resulted in the determination of 144 energy levels of the 4f4, 4f35d, 4f36s, and 4f36p configurations of Nd III, 105 of which were experimentally established for the first time. Of the 40 previously published Nd III levels, one was revised and 39 were confirmed.
Conclusions. The results will not only benchmark and improve future semi-empirical atomic structure calculations of Nd III, but also enable more reliable astrophysical applications of Nd III, such as abundance analyses of kilonovae and chemically peculiar stars, and studies of pulsational wave propagation in these stars.}, journal={Astronomy \& Astrophysics}, author={Ding, M. and Ryabtsev, A. N. and Kononov, E. Y. and Ryabchikova, T. and Clear, C. P. and Concepcion, F. and Pickering, J. C.}, year={2024}, month=apr, pages={A149}, language={en} }

@article{CarvajalGallego_Deprince_Palmeri_Quinet_2023, title={Expansion and line-binned opacities of samarium ions for the analysis of early kilonova emission from neutron star mergers}, volume={522}, ISSN={0035-8711}, DOI={10.1093/mnras/stad990},  number={1}, journal={Monthly Notices of the Royal Astronomical Society}, author={Carvajal Gallego, H and Deprince, J and Palmeri, P and Quinet, P}, year={2023}, month=june, pages={312–318} }

@article{Gallego_Deprince_Maison_Palmeri_Quinet_2024, title={Overview of the contributions from all lanthanide elements to kilonova opacity in the temperature range from 25 000 to 40 000 K}, volume={685}, rights={© The Authors 2024}, ISSN={0004-6361, 1432-0746}, DOI={10.1051/0004-6361/202347723},  journal={Astronomy \& Astrophysics}, publisher={EDP Sciences}, author={Carvajal Gallego, H and Deprince, J. and Maison, L. and Palmeri, P. and Quinet, P.}, year={2024}, month=may, pages={A91}, language={en} }

@article{CarvajalGallego_Deprince_Berengut_Palmeri_Quinet_2023, title={Opacity calculations in four to nine times ionized Pr, Nd, and Pm atoms for the spectral analysis of kilonovae}, volume={518}, ISSN={0035-8711}, DOI={10.1093/mnras/stac3129}, abstractNote={New atomic data for radiative transitions in Pr V–X, Nd V–X, and Pm V–X were determined by means of large-scale calculations involving three independent theoretical methods, i.e. the pseudo-relativistic Hartree–Fock method including core-polarization corrections (HFR+CPOL), the multiconfiguration Dirac–Hartree–Fock (MCDHF) method, and the configuration interaction many-body perturbation theoryMBPT) implemented in the ambit program. This multiplatform approach allowed us to estimate the reliability of the results obtained and to extract a large amount of energy levels, wavelengths, transition probabilities, and oscillator strengths for the determination of opacities required for the analysis of the spectra emitted in the early phases of kilonovae following neutron star mergers, i.e. for typical conditions corresponding to temperatures T &gt; 20000K, a density m−3, and a time after the merger t. Our radiative parameters were compared in detail with the few experimental data published so far and their impact on the calculated opacities, in terms of atomic computation strategy, was also examined.}, number={1}, journal={Monthly Notices of the Royal Astronomical Society}, author={Carvajal Gallego, H and Deprince, J and Berengut, J C and Palmeri, P and Quinet, P}, year={2023}, month=jan, pages={332–352} }

@article{Singh_Harman_Keitel_2025, title={Dielectronic recombination studies of ions relevant to kilonovae and non-LTE plasma}, url={http://arxiv.org/abs/2504.06639}, DOI={10.48550/arXiv.2504.06639}, abstractNote={This study presents calculations of rate coefficients, resonance strengths, and cross sections for the dielectronic recombination (DR) of Y^+, Sr^+, Te^2+, and Ce^2+--low-charge ions relevant to kilonovae and non-local thermodynamic equilibrium (non-LTE) plasmas. Using relativistic atomic structure methods, we computed DR rate coefficients under conditions typical of these environments. Our results highlight the critical role of low-lying DR resonances in shaping rate coefficients at kilonova temperatures (~ 10^4 K) and regulating charge-state distributions. Pronounced near-threshold DR resonances significantly influence the evolving ionization states and opacity of neutron star merger ejecta. Comparisons with previous studies emphasize the necessity of including high-n Rydberg states for accurate DR rate coefficients, especially for complex heavy ions with dense energy levels. Discrepancies with existing datasets underscore the need for refined computational techniques to minimize uncertainties. These results provide essential input for interpreting spectroscopic observations of neutron star mergers, including James Webb Space Telescope data. We also put forward suitable candidates for experimental studies, recognizing the challenges involved in such measurements. The data presented here have potential to refine models of heavy-element nucleosynthesis, enhance plasma simulation accuracy, and improve non-LTE plasma modeling in astrophysical and laboratory settings.}, note={arXiv:2504.06639 [astro-ph]}, number={arXiv:2504.06639}, publisher={arXiv}, author={Singh, Suvam and Harman, Zoltán and Keitel, Christoph H.}, year={2025}, month=apr }

@article{FerreiradaSilva2026, title={Resonant Excitation in r-Process Collision Strengths for Non-LTE Kilonova Modelling}, volume={13}, rights={http://creativecommons.org/licenses/by/3.0/}, ISSN={2673-9984}, DOI={10.3390/psf2026013005}, number={1}, journal={Physical Sciences Forum}, publisher={Multidisciplinary Digital Publishing Institute}, author={Ferreira da Silva, Ricardo and Leitão, Luis and Flörs, Andreas and Campante, Tomás and Garcia, Daniel and Sampaio, Jorge and Martínez-Pinedo, Gabriel and Pires Marques, José}, year={2026}, pages={5}, language={en} }

@article{Deprince_Maison_2026, title={Benchmarking the Plane-Wave Born and Distorted Waves approximations for electron-impact collision strength computations: the sample case of Sr II}, volume={21}, ISSN={1880-6821}, DOI={10.1585/pfr.21.2401005}, journal={Plasma and Fusion Research}, publisher={The Japan Society of Plasma Science and Nuclear Fusion Research (JSPF)}, author={Deprince, Jerome and Maison, Lucas}, year={2026}, language={Anglais} }
\bibliographystyle{aasjournal}
\end{document}